\documentclass[manuscript,screen]{acmart}
\AtBeginDocument{%
  }

\setcopyright{acmlicensed}
\copyrightyear{2018}
\acmYear{2018}
\acmDOI{XXXXXXX.XXXXXXX}
\acmConference[Conference acronym 'XX]{Make sure to enter the correct
  conference title from your rights confirmation email}{June 03--05,
  2018}{Woodstock, NY}
\acmISBN{978-1-4503-XXXX-X/2018/06}

\usepackage{booktabs}
\usepackage{multirow}
\usepackage{tabularx}
\usepackage{listings}
\usepackage{xcolor}
\usepackage{wasysym}
\usepackage{threeparttable}
\usepackage{listings}
\usepackage{xcolor} 

\usepackage[most]{tcolorbox}

\tcbset{
  findingstyle/.style={
    enhanced,
    breakable,
    colback=black!2,
    colframe=black!40,
    boxrule=0.4pt,
    arc=1mm,
    left=4pt,right=4pt,top=3pt,bottom=3pt,
  }
}

\newtcolorbox{findingbox}[1][]{findingstyle,#1}
\usepackage[table]{xcolor} 
\usepackage{makecell}
\usepackage[normalem]{ulem}

\lstdefinestyle{promptstyle}{
  basicstyle=\ttfamily,
  breaklines=true,
  breakatwhitespace=true,
  columns=fullflexible,
  keepspaces=true,
  showstringspaces=false,
  upquote=true,
  frame=single,
  rulecolor=\color{black},
  xleftmargin=0.6em,
  xrightmargin=0.6em
}

\begin{document}

\title{Can Language Models Go Beyond Coding? \\ Assessing the Capability of Language Models to Build Real-World Systems}

\author{Chenyu Zhao}
\email{zhaochenyu@mail.nankai.edu.cn}
\affiliation{%
  \institution{Nankai University}
  \city{Tianjin}
  \country{China}
}

\author{Shenglin Zhang}
\authornote{Corresponding author.}
\email{zhangsl@nankai.edu.cn}
\affiliation{%
  \institution{Nankai University}
  \city{Tianjin}
  \country{China}
}
\additionalaffiliation{%
  \institution{Haihe Laboratory of Information Technology Application Innovation}
  \city{Tianjin}
  \country{China}
}
\additionalaffiliation{%
  \institution{Key Laboratory of Data and Intelligent System Security, Ministry of Education}
  \city{Tianjin}
  \country{China}
}

\author{Zeshun Huang}
\email{2213900@mail.nankai.edu.cn}
\affiliation{%
  \institution{Nankai University}
  \city{Tianjin}
  \country{China}
}

\author{Weilin Jin}
\email{2401112012@stu.pku.edu.cn}
\affiliation{%
  \department{School of Computer Science}
  \institution{Peking University}
  \city{Beijing}
  \country{China}
}

\author{Yongqian Sun}
\email{sunyongqian@nankai.edu.cn}
\affiliation{%
  \institution{Nankai University}
  \city{Tianjin}
  \country{China}
}
\additionalaffiliation{%
  \institution{Tianjin Key Laboratory of Software Experience and Human Computer Interaction}
  \city{Tianjin}
  \country{China}
}

\author{Dan Pei}
\email{peidan@tsinghua.edu.cn}
\affiliation{%
  \institution{Tsinghua University}
  \city{Beijing}
  \country{China}
}

\author{Chaoyun Zhang}
\email{chaoyun.zhang@microsoft.com}
\affiliation{%
  \institution{Microsoft}
  \city{Beijing}
  \country{China}
}

\author{Qingwei Lin}
\email{qlin@microsoft.com}
\affiliation{%
  \institution{Microsoft}
  \city{Beijing}
  \country{China}
}

\author{Chetan Bansal}
\email{chetanb@microsoft.com}
\affiliation{%
  \institution{Microsoft}
  \city{Redmond}
  \country{USA}
}

\author{Saravan Rajmohan}
\email{saravan.rajmohan@microsoft.com}
\affiliation{%
  \institution{Microsoft}
  \city{Redmond}
  \country{USA}
}

\author{Minghua Ma}
\email{minghuama@microsoft.com}
\affiliation{%
  \institution{Microsoft}
  \city{Seattle}
  \country{USA}
}

\renewcommand{\shortauthors}{Trovato et al.}
\newcommand{\mh}[1]{{\color{blue}[MH: #1]}}
\newcommand{\ie}{\textit{i.e.,}~}
\newcommand{\eg}{\textit{e.g.,}~}
\newcommand{\etal}{\textit{et al.}~}
\def\name{\textit{Build-bench}}

\begin{abstract}

Large language models (LLMs) have shown growing potential in software engineering, yet few benchmarks evaluate their ability to repair software during migration across instruction set architectures (ISAs).
Cross-ISA migration, such as between \texttt{x86\_64} and \texttt{aarch64}, requires handling complex dependencies, heterogeneous toolchains, and long build logs while ensuring executable verification.
To address this challenge, we present \name{}~\footnote{Homepage of \name{}: \url{https://aiops-lab-nku.github.io/Build-bench/}, Code at \url{https://github.com/zcyyc/Build-bench}}, an end-to-end benchmark that systematically evaluates the capability of LLMs to repair build failures in cross-ISA settings.
\name{} collects 268 real-world failed packages and integrates auxiliary tools including \textit{Structure Extraction}, \textit{File Content Extraction}, \textit{Content Modification}, and \textit{Build Verification} to support autonomous, tool-augmented reasoning.
The repair process operates in an iterative loop where, upon failure, the model receives updated build logs and previous repair outcomes to refine subsequent attempts.
Through a comparative evaluation across the studied models, \name{} reveals that current models achieve a maximum build success rate of 63.19\% and tool usage patterns differ significantly across models.
By coupling real build environments with verifiable outcomes, \name{} establishes the first architecture-aware benchmark for studying LLM-based software build and repair.

\end{abstract}



\received{20 February 2007}
\received[revised]{12 March 2009}
\received[accepted]{5 June 2009}

\maketitle

\section{Introduction}

Large language models (LLMs) have become increasingly integrated into modern software development ecosystems, driving remarkable progress in the automation of diverse software engineering tasks.
Within the field of software engineering, these LLMs now assist with code generation~\cite{DBLP:conf/icse/JiangDTLJ025}, test synthesis~\cite{DBLP:conf/icse/LemieuxILS23}, project debugging~\cite{DBLP:conf/icse/HuangWLW0CHW24}, error detection and issue resolution~\cite{DBLP:conf/dsa/KangAL24}, which together substantially improve developer productivity~\cite{DBLP:journals/corr/abs-2406-00515}.

To systematically evaluate the strengths and limitations of LLMs in programming contexts, researchers have developed a series of benchmarks that target different aspects of software intelligence.
CoderEval~\cite{DBLP:conf/icse/YuSRZZMLLWX24} focuses on functional code generation,  OpenRCA~\cite{DBLP:conf/iclr/XuZZHZLPHZ025} benchmarks LLMs for intelligent root-cause analysis in real operational environments, and SWE-bench~\cite{DBLP:conf/iclr/JimenezYWYPPN24} assesses end-to-end issue resolution on real-world repositories.
Recent extensions such as SWE-bench-Live~\cite{DBLP:journals/corr/abs-2505-23419} further enable dynamic evaluation on continuously updated repositories, providing deeper insight into the robustness of autonomous code repair systems.
However, these benchmarks primarily focus on functionally-defined repair tasks under \textbf{homogeneous software and hardware environments.} 
They assume that code execution semantics remain consistent across platforms, and that repair success can be verified through test cases or oracle assertions. 
In contrast, real-world software ecosystems increasingly span \textbf{heterogeneous computing architectures}, introducing fundamental differences that challenge this assumption.


Driven by the demand for energy-efficient, cloud-native, and heterogeneous systems, the global computing landscape is undergoing a large-scale transition in instruction set architectures (ISAs)~\cite{DBLP:conf/asplos/XingXWSRBB25}.
Among the major ISAs, \textbf{x86\_64} and \textbf{aarch64 (ARM64)} dominate modern computing yet differ substantially in register organization, memory models, compiler behaviors, and toolchains. 
This divergence has led to widespread cross-ISA migration efforts.
Apple’s transition to ARM-based M-series chips~\cite{apple_styleguide_2024}, Amazon’s deployment of Graviton processors~\cite{AWSGraviton2023}, and Microsoft’s ARM-compatible Windows platforms exemplify this shift. 
Ensuring that large-scale open-source software ecosystems remain portable and buildable across such heterogeneous environments has thus become an urgent challenge for sustainable software evolution. 

To maintain correctness and portability during migration between x86\_64 and aarch64 platforms, large software ecosystems (\eg operating systems, middleware, and package repositories) require extensive source-level refactoring and repair.
Despite several industrial efforts toward cross-ISA migration~\cite{AlibabaYitianBlog2021,apple_styleguide_2024,AWSGraviton2023}, they rely primarily on internal and ad-hoc evaluation processes, which hinder consistent performance comparison and reproducibility across systems.
A major obstacle to automating this process lies in the inherent heterogeneity of cross-ISA build failures, which span multiple stages from environment configuration to packaging (as analyzed in Table~\ref{tab:build-failure-classification}). 
Although prior work has explored build repair, diagnosis, and feedback-driven recovery, existing approaches typically provide only localized or stage-specific support, such as history-driven script repair, dependency-oriented diagnosis, or static root-cause prediction~\cite{DBLP:conf/icse/HassanW18, DBLP:journals/tosem/NourryKSSK25, DBLP:conf/issta/FanWW0SZ20, DBLP:conf/kbse/Zhang0H0Z22, DBLP:conf/sigsoft/Zhang0C0Z19}. 
Such methods struggle with the non-linear nature of migration failures, where the effective repair sequence (\eg coordinating structure extraction, dependency adjustment, and macro modification) cannot be fully pre-defined. 
Addressing these failures requires an orchestrator capable of dynamically navigating the tool space and synthesizing heterogeneous feedback from an executable build environment. 
However, the software engineering community still lacks a benchmark for systematically evaluating whether large language models (LLMs) can understand, adapt, and repair software packages across heterogeneous ISAs.
To bridge this gap, we propose \name{}, the first benchmark designed to evaluate whether LLMs can interpret build-failure contexts, generate effective repairs, and achieve successful rebuild during cross-ISA software packages migration.
Constructing such a benchmark introduces multiple challenges.

\begin{figure}[h]
  \centering
  \includegraphics[width=\linewidth]{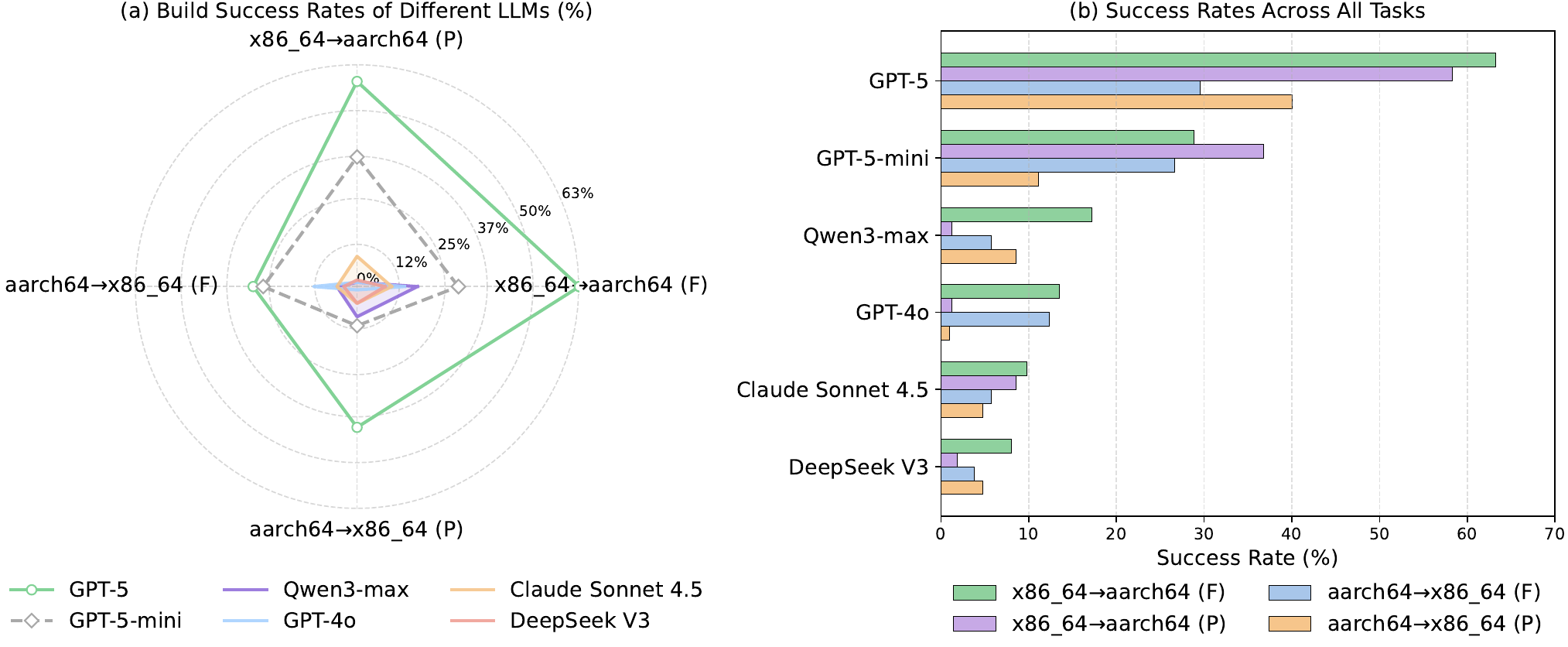}
  \caption{Comparison of different large language models (LLMs) in cross-ISA build repair tasks. 
    (a) shows the success rates (\%) achieved on four migration scenarios (\texttt{x86\_64$\rightarrow$aarch64 (F)}, \texttt{x86\_64$\rightarrow$aarch64 (P)}, \texttt{aarch64$\rightarrow$x86\_64 (F)}, and \texttt{aarch64$\rightarrow$x86\_64 (P)}), where F denotes \textit{Full File Generation} and P denotes \textit{Patch Generation}.
    (b) summarizes the overall success rates across all tasks for each model.}
    \label{fig:radar-llm-comparison}
\end{figure}

\begin{itemize}
\item \textbf{Challenge 1: Capturing the complex reasoning context underlying multi-layer builds.}  
Cross-ISA build failures rarely occur within a single source file; instead, they arise from intricate dependencies among specification descriptors that define metadata, dependencies, and architecture-specific conditions, together with build scripts, compiler options, and heterogeneous toolchains.
Existing benchmarks~\cite{DBLP:conf/iclr/JimenezYWYPPN24, DBLP:journals/corr/abs-2505-23419, zhang2025buildbench} focus on single-architecture settings and typically provide only file-level inputs with test-based oracles, which are insufficient for evaluating reasoning over such system-level interactions.

\item \textbf{Challenge 2: Addressing large-scale codebases and complex environments in cross-ISA migration.}  
On average, each package contains 366 files and over 55,000 lines of code, spanning multiple programming languages (\eg C/C++, Rust, Python, JavaScript), shell scripts, Makefiles, and architecture-specific configuration files.
During cross-ISA migration, the model must not only modify the relevant components but also maintain global consistency across source files, dependency declarations, compiler toolchains, and environment variables.
Such large-scale, multi-language, and architecture-sensitive characteristics fundamentally distinguish this task from small-scale, single-file program repair benchmarks.

\item \textbf{Challenge 3: Achieving verifiable, end-to-end evaluation.}  
Most existing benchmarks evaluate LLMs in a single-turn setting, where one repair attempt determines success.  
In contrast, cross-ISA software package migration relies heavily on iterative feedback.
Build logs provide valuable contextual signals that reveal deficiencies in prior modifications and guide progressive refinement.  
A robust benchmark must assess whether an LLM can leverage such iterative feedback to continuously improve repair accuracy.  
Moreover, migration success must be validated through executable, end-to-end rebuild rather than textual comparison or test passing.  
A repair is considered successful only when the package can be fully rebuilt under the target architecture within a controlled, reproducible environment.
Consequently, constructing a comprehensive benchmark that integrates iterative reasoning capabilities with verifiable system execution remains a non-trivial challenge.
\end{itemize}

To address these challenges, \name{} implements an end-to-end evaluation pipeline that integrates automated build verification with iterative reasoning.
Following prior work that employs standardized orchestration frameworks such as the Model Context Protocol (MCP)~\cite{hou2025modelcontextprotocolmcp}, \name{} takes failed build packages (\ie real packages that succeed on one ISA but fail on another) as input and orchestrates multiple external tools, including \textit{Structure Extraction}, \textit{File Content Extraction}, and source archive \textit{Compression/Decompression}, for package analysis and source manipulation.
In the validation phase, \name{} leverages the \textbf{Open Build Service (OBS)} \footnote{\url{https://openbuildservice.org/}}, which provides a reproducible environment to enable online package building. 
After the LLM analyzes the failure and proposes minimal modifications based on the system prompt, the repaired software package is automatically uploaded to OBS for rebuilding. 
\name{} then invokes the \textit{Check Build Result} tool to retrieve build results from OBS.
If the build fails, the updated logs and diffs are returned to the LLM for further repair attempts; if it succeeds or the maximum iteration limit is reached, the repair task terminates. 
This multi-round repair workflow mirrors real industrial migration practices.
Through this design, \name{} enables the first systematic evaluation of LLMs on repairing build failures during cross-ISA software migration.

\name{} consists of 268 real-world software packages, where 163 fail on aarch64 but succeed on x86\_64, and 105 in the reverse direction, forming a realistic and challenging corpus of cross-ISA migration failures.
We evaluate six state-of-the-art models, namely \textit{GPT-5}~\cite{OpenAI2025IntroducingGPT5}, \textit{GPT-5-mini}~\cite{OpenAI2025IntroducingGPT5}, \textit{GPT-4o}~\cite{OpenAI2025IntroducingGPT4o}, \textit{Claude Sonnet 4.5}~\cite{Claude2025Sonnet}, \textit{DeepSeek V3}~\cite{DeepSeekV3TechReport}, and \textit{Qwen3-max}~\cite{Alibaba2025Qwen3Max}, under both \textit{Full File Generation} and \textit{Patch Generation} repair modes.
In the \textit{Full File Generation} mode, once an LLM identifies a file containing errors, it directly generates a complete revised version that replaces the original file.
In the \textit{Patch Generation} mode, the LLM specifies fine-grained edits such as additions, deletions, or modifications, and the auxiliary tools automatically apply these changes to the source tree.

Fig.~\ref{fig:radar-llm-comparison} summarizes the success rates of all models across two migration directions and two repair strategies.
As shown in Fig.~\ref{fig:radar-llm-comparison} (a), GPT-5 maintains consistently high success rates across all scenarios, reaching 63.19\% on the x86\_64 $\rightarrow$ aarch64 by \textit{Full File Generation} in particular.
Fig.~\ref{fig:radar-llm-comparison} (b) further aggregates the overall success rates across all tasks, revealing a clear performance hierarchy among models.
Although GPT-5 and GPT-5-mini nearly outperform the others, large language models still struggle with large-scale, heterogeneous, and architecture-specific repair tasks.
By integrating realistic build feedback and executable verification, \name{} provides a unified foundation for evaluating and enhancing software package repair across multiple ISAs.

In summary, the main contributions of \name{} are as follows:
\begin{itemize}
    \item \textbf{A new benchmark and corpus for cross-ISA build.}
We present \textbf{\name{}}, the first benchmark designed to evaluate large language models (LLMs) in repairing software packages that fail during migration across heterogeneous instruction set architectures (ISAs).
The benchmark includes 268 real-world software packages, providing a realistic and reproducible corpus for studying architecture-aware reasoning.
\item \textbf{An end-to-end evaluation framework with iterative build verification.}
\name{} integrates real build environments and automated verification pipelines, enabling executable evaluation through multiple repair iterations.
The framework captures both build logs and prior modifications as feedback, allowing systematic assessment of LLMs’ ability to refine repair strategies under dynamic build contexts.
\item \textbf{Comprehensive empirical analysis and quantitative insights.}
We evaluate six state-of-the-art LLMs covering both proprietary
and open-source (GPT-5, GPT-5-mini, GPT-4o, Claude Sonnet 4.5, DeepSeek V3, and Qwen3-max) across \textit{Full File Generation} and \textit{Patch Generation} modes.
Our results, summarized in Fig.~\ref{fig:radar-llm-comparison}, reveal significant performance variance among models, expose persistent weaknesses in multi-file reasoning and architecture-specific adaptation, and establish baselines for future research on cross-ISA software repair.
\end{itemize}

\section{Background}
\subsection{Package Building}
Software package building~\cite{10.5555/1971982} is a critical process in software engineering that automates the transformation of source code, specification files (including metadata, dependencies, and architecture-specific conditions), and build scripts into distributable binary packages. 
It is widely used in operating system distributions (\eg openSUSE, Debian, Fedora)~\cite{opensuse2022,debian2018,fedora2022} and large-scale software ecosystems’ continuous integration (CI) pipelines.
As software systems continue to grow in scale, package building faces increasing challenges~\cite{northrop2006ultra,mens2024overview,moreau2023containers} related to dependency complexity, environmental inconsistencies, and reproducibility. 
To address these issues, researchers have explored reproducible builds and automated dependency management mechanisms to improve build stability and security~\cite{pashchenko2020qualitative}.

In practice, the Open Build Service (OBS), an open-source distributed build and release platform, has been widely adopted. 
OBS supports multi-architecture builds, automated dependency resolution, isolated build environments, and version tracking, enabling the generation of installable packages for multiple distributions from a single platform and providing scalable support for reproducible builds and continuous delivery. 
Nevertheless, build failures remain common~\cite{lou2020understanding}, often caused by missing dependencies, version conflicts, or errors in build scripts. 
As the scale of open-source projects increases, build logs~\cite{he2021survey,brandt2020logchunks} become larger and more complex, making manual analysis inefficient and error-prone.

To address these challenges, both academia and industry have proposed various approaches for diagnosing and repairing build failures, including log pattern mining~\cite{brandt2020logchunks,bushong2020matching} and history-based automated repair~\cite{parashar2017package}. 
Recently, artificial intelligence and large language models~\cite{zhang2026systematicliteraturereviewlarge,zhang2024survey,belzner2023large,jin2026benchmark, Liu2025_OpsEval} (LLMs) have shown promising potential for understanding build logs and performing automated repairs. 
LLM-based methods~\cite{yang2025survey,cxxcrafter,zhong2025logupdater,zhang2024failure,yu2024monitorassistant,wang2024large,chen2025aiopslab} can identify error patterns, locate root causes, and generate repair suggestions from complex build logs, offering new opportunities for intelligent and automated software package building.

In summary, while OBS provides a reliable infrastructure for large-scale package building, build failures and their repair remain major obstacles to efficiency. Leveraging intelligent techniques~\cite{shethiya2024engineering,hou2024large,shethiya2024engineering} such as LLMs for automatic analysis and repair of build logs has thus emerged as a key research direction in software engineering automation.

\subsection{Cross-instruction Set Architecture Migration}
\subsubsection{The Concept of ISA}
The Instruction Set Architecture (ISA)~\cite{barbacci2012instruction,mokhov2013synthesis} serves as the abstract model of a computer, defining the set of commands that the processor can execute, including data types, registers, and memory addressing modes. 
It forms the crucial interface between hardware and software. 
Among the diverse ISAs, x86\_64 and aarch64 (ARM64)~\cite{gupta2021changing,sankaralingam2013detailed} have been the subjects of extensive research and deployment, dominating the server and mobile computing landscapes, respectively. 
The distinctions among ISAs, such as their instruction sets, calling conventions, and memory models, are fundamental sources of challenges in cross-ISA software migration~\cite{bapat2023hetmigrate}.
\subsubsection{Cross-ISA Migration}
With the rapid development of heterogeneous computing environments~\cite{de2019software}, software package building is no longer limited to a single hardware architecture. 
Cross-ISA migration aims to ensure that software packages can be correctly built and executed on different architectures~\cite{ford2022cost,ford2021migrating} (\eg x86\_64, aarch64), thereby supporting multi-platform deployment and performance optimization. 
However, due to differences in instruction sets~\cite{ford2021migrating}, compilation toolchains~\cite{ketata2015performance}, and dependency ecosystems~\cite{decan2019empirical}, migration often faces significant challenges, including compilation failures, dependency conflicts, and runtime errors~\cite{rausch2017empirical,tellnes2013dependencies}.

In open-source ecosystems, comprehensive multi-architecture support is still limited~\cite{sowinski2024overview}. 
Developers often need to manually modify build scripts or source code to adapt to the target architecture. 
Such manual intervention not only increases maintenance costs but also introduces potential errors, leading to frequent build failures during migration. 

Existing research has highlighted the risks associated with hardware-specific dependencies during architecture migrations. 
For instance, Ford et al. and Davi et al.~\cite{ford2021migrating,davi2013gadge} point out that as servers and mobile devices transition to the ARM architecture, packages reliant on x86-specific features like SSE instructions or particular byte ordering are prone to build failures. 
Similarly, Wressnegger et al.~\cite{wressnegger2016twice} demonstrate in their study ``Twice the Bits, Twice the Trouble'' that migrating from 32-bit to 64-bit environments introduces pointer size changes, which can affect memory alignment or cause integer overflows, thereby compromising build correctness and runtime behavior.
Due to inconsistent dependencies~\cite{cataldo2009software}, environment configurations~\cite{decan2019empirical}, or architecture-specific compilation flags, many packages still fail to build successfully. 
Therefore, enhancing the automation and intelligence of cross-ISA package migration is crucial for reducing maintenance costs, improving build success rates, and supporting the sustainable development of large-scale software ecosystems, making it a key research direction in software engineering and system maintenance~\cite{dittrich2014software,northrop2006ultra,venters2018software}.

\section{\name}
In this section, we introduce \name{}, a benchmark designed to evaluate large language models (LLMs) on cross-instruction set architecture (cross-ISA) repair and build tasks. 
Unlike prior benchmarks that focus on source-level bug fixing or single-architecture builds, \name{} centers on real-world software packages that fail during migration between x86\_64 and aarch64. 
It aims to assess whether LLMs can autonomously diagnose build failures, apply targeted code or configuration modifications, and verify the repaired packages through executable rebuild on a real build service.  

\subsection{Overview}
\label{sub: overview}
\begin{figure}[t]
  \centering
  \includegraphics[width=\linewidth]{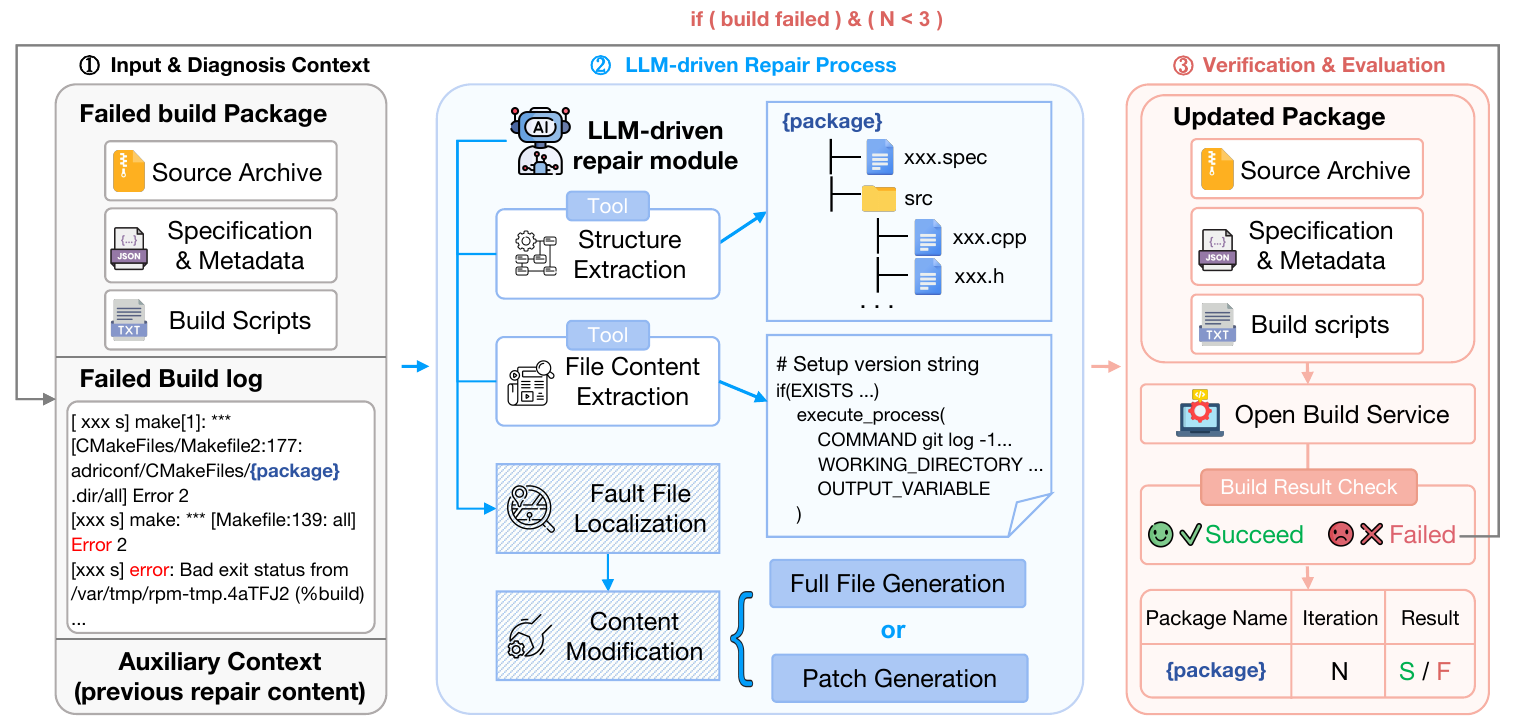}
  \caption{The automatic cross-ISA repair and build pipeline of \name{}. If the build fails and the maximum iteration $N_{\max}=3$ is not reached, the process repeats with the updated build log as well as the previous repair content.}
    \label{fig:framework}
\end{figure}

\begin{table*}[t]
  \centering
  \caption{Classification of Build Failure Cases. ``Category'' refers to the expert-defined failure category, ``Subcategory'' is the further division based on the failure, and ``Count'' indicates the number of packages in each sub-category.}
  \label{tab:build-failure-classification}
  \newcolumntype{L}[1]{>{\raggedright\arraybackslash}p{#1}}
    \begin{tabularx}{\textwidth}{
      L{0.18\textwidth}
      X
      c
      X
    }

    \toprule
    \textbf{Category} & \textbf{Subcategory} & \textbf{Count} & \textbf{Description} \\
    \midrule
    \multirow{5}{0.18\textwidth}{Build Preparation Error}
      & Environment and Dependency Issues  & 34 & Missing macros, incompatible toolchains, or unresolved dependencies preventing configuration. \\
      & Compiler and Flag Configuration Errors  & 44 & Invalid compiler flags, toolchain incompatibilities, or duplicate arguments. \\
      \cmidrule{2-4}
      & \textbf{Sum}           & \textbf{78} & \\
    \midrule
    \multirow{7}{*}{Compilation Error}
      & Build System and Compiler Configuration Failures  & 44 & Build script, compiler flag, or linker-level incompatibility. \\
      & Compiler and Type System Errors  & 57 & Type mismatch, prototype conflict, missing headers, language standard incompatibility. \\
      & Warning Escalation and Policy-Induced Failures  &  15  & Warnings promoted to errors due to strict compiler policies. \\
      \cmidrule{2-4}
      & \textbf{Sum}         & \textbf{116} & \\
    \midrule
    \multirow{8}{*}{Packaging Error}
      & Missing or Unpackaged Artifacts   & 14 & Missing or unreferenced build outputs such as binaries, manpages, or documentation. \\
      & RPM Script and Build Step Failures  & 7 & Non-zero exit during \%build, \%install, or Makefile targets. \\
      & Specification or Policy Violations  & 3 & Packaging policy issues such as duplicate installs or missing validation sections. \\
      \cmidrule{2-4}
      & \textbf{Sum}                      & \textbf{24} & \\
    \midrule
    \multirow{6}{*}{Test Failure}
      & Functional and Assertion Failures   & 20 & Core logic or assertion mismatches during testing. \\
      & Environment Setup Failures  & 10 & Missing or incompatible test dependencies or restricted runtime environments. \\
      & Runtime and Execution Failures  & 12 & Resource, crash, or timeout errors during test execution. \\
      \cmidrule{2-4}
      & \textbf{Sum}                      & \textbf{42} & \\
    \midrule
    \multirow{3}{0.18\textwidth}{Environment/\\Infrastructure Error}
      & Host or Virtualization Failure    &  8 & Build interrupted due to VM shutdown, power loss, or infrastructure termination. \\
      \cmidrule{2-4}
      & \textbf{Sum}                      & \textbf{8} & \\
    \midrule
    \textbf{Total} &  & \textbf{268} & \\
    \bottomrule
  \end{tabularx}
\end{table*}

The overall workflow of \name{}, illustrated in Fig.~\ref{fig:framework}, consists of three major stages:
(1)~\textit{Input \& Diagnosis Context},
(2)~\textit{LLM-driven Repair Process}, and
(3)~\textit{Verification \& Evaluation}.
In the first stage, \name{} collects essential contextual artifacts from each failed package directory to support diagnosis and repair.
Specifically, the inputs include:
\begin{itemize}
  \item \textbf{Source archives} (\eg \texttt{.tar.gz}, \texttt{.bz2}), which contain the original code base and associated assets.  
  \item \textbf{Specification and metadata} (\eg \texttt{.spec}, \texttt{.changes}), which define build configurations, dependency declarations, and version updates.  
  \item \textbf{Build scripts} (\eg \texttt{.service}, \texttt{.desktop}), which provide configuration or service-level integration scripts.  
  \item \textbf{Failed build logs}, which records the compiler output and error traces during the failed build.
\end{itemize}

These inputs and diagnosis contexts provide the diagnostic foundation for failure analysis.
For iterations beyond the first, \name{} enriches the inputs with the latest package state, the updated build log, and an \textbf{auxiliary context} that preserves modifications from previous attempts, thereby allowing the model to reason with historical information.

The second stage leverages an LLM-driven repair module based on the Model Context Protocol (MCP), which allows dynamic interaction between the model and a suite of external tools for information extraction and content modification.
To assess how different editing granularities affect model performance, the repair process produces two experimental variants evaluated in Section~\ref{subsec: impact of repair strategy}:
\textit{Full File Generation}, where the model regenerates the entire faulty file while preserving its structure and minimal edits, and
\textit{Patch Generation}, where explicit line-level modifications are output in a diff-like format automatically applied on the relevant file by \name{}.

Finally, the updated package is rebuilt on the Open Build Service (OBS) to verify whether the repair succeeds.
If the rebuild fails and the maximum iteration threshold is not reached, the process repeats with the updated inputs.
This iterative workflow enables reproducible, end-to-end evaluation of LLM-based repair performance across heterogeneous ISAs.

\subsection{Benchmark Construction}
We begin by collecting software packages from the official OBS repositories, where 17,001 packages successfully build on x86\_64 and 16,892 packages successfully build on aarch64.
To construct a representative yet computationally manageable benchmark, we randomly sample a subset of successfully built packages from each architecture as the source sets.
Each sampled package is replicated into a controlled workspace and rebuilt on the opposite ISA.
Packages that fail to compile under the new architecture are retained as the initial migration-failure candidates.

To ensure data quality and reproducibility, each failed package is rebuilt on OBS again to confirm that the failure can be consistently reproduced.
We then remove incomplete or corrupted packages whose source archives or specification files are missing.
After filtering and characterization, \name{} contains \textbf{163 packages} that build successfully on x86\_64 but fail on aarch64, and \textbf{105 packages} that build successfully on aarch64 but fail on x86\_64, forming a corpus of 268 reproducible cross-ISA build failures.

To better understand the composition of the corpus, we analyze the build logs of failed packages using LLM-assisted summarization based on \textit{GPT-5-mini}.
To improve transparency and reproducibility, we conducted a full manual validation of the failure categorization. 
Three experienced software engineers independently reviewed the build logs and the LLM-assisted category assignments for all packages in the corpus. 
After discussion and consensus, the human-validated labels were treated as the reference categorization. 
We found that 95.90\% of the LLM-assisted labels were consistent with the final human consensus.
The few misclassified cases were subsequently corrected based on the engineers' consensus, and the updated labels were used to produce the final statistics. 
Note that these categories are used solely for descriptive analysis of failure diversity and do not affect any evaluation metrics in the benchmark.

As summarized in Table~\ref{tab:build-failure-classification}, the failures are categorized into five major types: \textit{Build Preparation Error}, \textit{Compilation Error}, \textit{Packaging Error}, \textit{Test Failure}, and \textit{Environment/Infrastructure Error}.
Specifically, build preparation errors arise from missing dependencies, misconfigured toolchains, or invalid compiler flags; compilation errors stem from language or build-system incompatibilities; packaging errors involve incomplete or mis-specified build artifacts;  test failures reflect runtime or functional regressions observed during verification; and environment/infrastructure errors correspond to external interruptions such as unexpected VM shutdowns.

These failure patterns collectively span the entire build process, covering multiple stages from environment setup and dependency resolution to compilation, packaging, and runtime validation.
Moreover, the failures involve heterogeneous sources of information, such as source archive, specification and metadata, build scripts, and test environments.
Such observations confirm that cross-ISA software package migration is inherently complex and multi-dimensional, requiring reasoning across multiple build stages, heterogeneous artifacts, and execution contexts.
The diversity and realism of these failures provide a credible foundation for constructing a meaningful benchmark.

\subsection{LLM-driven repair process}
The LLM-driven repair module constitutes the core intelligence of \name{}, where a LLM orchestrates multiple specialized tools.
Unlike static pipelines that follow fixed procedures, this module enables dynamic, context-aware interaction.
It transforms the repair process into a reasoning-driven workflow, where tools serve as external functional modules, enabling direct evaluation of the LLM’s capability in tool orchestration and complex problem solving.

\subsubsection{Tool Ecosystem and Orchestration via MCP}
To strictly decouple LLM reasoning from the underlying functional tool suite and enable fair, reproducible evaluation across models, \name{} adopts the MCP~\cite{hou2025modelcontextprotocolmcp} as a standardized interface.
Under MCP, each auxiliary tool is registered with a unified schema specifying its name, purpose, parameters, and expected outputs.
Importantly, MCP itself does not store memory nor perform any reasoning in \name{}.
Instead, it serves as a integration layer that exposes a consistent tool invocation interface and a shared, transparent action space.
This design ensures that all evaluated LLMs interact with the same tool definitions and operational semantics, allowing us to compare their tool orchestration capabilities without confounding factors from model-specific tool wrappers or tightly coupled controller logic.
During the repair process, the LLM dynamically discovers and invokes these tools at runtime and integrates the returned information into its evolving repair hypothesis. 
All interactions are carefully logged with timestamps and specific return results, ensuring transparent and reproducible orchestration.

To provide contextual knowledge of the failed package, \name{} provides the \textit{Structure Extraction} and \textit{File Content Extraction} tools.  
\begin{itemize}
    \item \textit{Structure Extraction Tool} constructs a hierarchical representation of the package repository by recursively scanning the source tree and excluding non-essential documentation and metadata files (\eg \texttt{README.md}, \texttt{LICENSE}, \texttt{doc/}).
    For each subdirectory, it records the relative structure of source archive, specification files, and build scripts, and generates a compact JSON-like schema. 
    Furthermore, it leverages language-aware parsers such as \texttt{tree-sitter}~\cite{tree-sitter} to recognize programming constructs in Python, C/C++, Java, Rust, Go, and TypeScript files, extracting class and function definitions together with line spans and method-level boundaries.
    \item \textit{File Content Extraction Tool} retrieves the full textual content of target files for detailed reasoning, ensuring that the LLM operates on complete rather than truncated contexts.
\end{itemize}

Guided by the outputs of these auxiliary tools, the LLM performs autonomous reasoning and infers which file or fragments are most likely responsible for the current build failure. 
Once the target files are identified, the model determines an appropriate repair strategy based on the prompt, as detailed in Section~\ref{sub: prompt design}.
To operationalize these repair decisions, \name{} provides a \textit{Content Modification Tool} that automatically applies the model’s edits to the corresponding files.  

\subsubsection{Iterative Reasoning and Verification Loop}
\label{subsubsec: iterative reasoning}
The repair process in \name{} is not a single-turn interaction but an iterative reasoning loop that incorporates build feedback into subsequent repair attempts.
Within each iteration, the LLM autonomously performs multiple rounds ($T$) of tool invocations, including verification by uploading the modified package to the Open Build Service (OBS).
For any iteration ($N > 1$), the input prompt is programmatically enriched with three specific components: 
(1) the concrete file modifications applied in the prior iteration together with their updated content, 
(2) the updated build log returned by the OBS, and 
(3) the latest software package repository. 
These artifacts are explicitly reintroduced into the prompt as observable execution feedback. 

To mitigate procedural redundancies, preserve computational tractability, and balance exploratory reasoning depth with operational overhead, the iterative framework is governed by two unified hyperparameters: the per-iteration tool-call limit ($T_{\max}=20$) and the maximum number of repair iterations ($N_{\max}=3$).
All tool invocations, including package uploads and build-result retrieval, are counted toward this per-iteration limit. 
Multiple uploads may occur within a single iteration, and such redundant operations consume additional tokens and wall-clock time.
These costs are therefore reflected in the reported efficiency and token-utilization metrics, as they form part of the \name{}’s evaluation of procedural efficiency and tool orchestration behavior. 
A detailed empirical justification for these settings, including an analysis of the marginal gains across iterations, is provided in Section~\ref{subsubsec: analysis of tool calls}.

An iteration terminates when the LLM’s response no longer indicates a tool call or when the invocation limit $T_{\max}$ is reached.
Importantly, the system does not forcibly terminate an iteration upon a build failure. 
Instead, iteration boundaries are primarily model-driven rather than imposed by the interface. 
If the package remains unsuccessful when an iteration ends, \name{} proceeds to the next iteration (provided $N < N_{\max}$) by re-synchronizing the prompt with the latest repository state and updated build log.
Through this iterative design, the LLM autonomously performs progressive failures correction, invokes external tools as needed, and proposes improved modifications to resolve remaining issues.  
Such an iterative repair loop enables progressive reasoning under real feedback, allowing the benchmark to assess not only the model’s static repair capability but also its ability to adapt, reflect, and converge toward a successful cross-ISA build.

\subsection{Prompt Design}
\label{sub: prompt design}

To improve transparency and reproducibility, we describe the prompt structure at a modular level. 
Each prompt in \name{} consists of four primary modules:
(1) an \emph{Instruction Module} specifying the repair objectives and guiding principles such as minimalism, completeness, and style preservation;
(2) a \emph{Context Module} providing hierarchical package structure, build logs, and other failure context;
(3) a \emph{Tool Interface Module} defining available tool schemas and execution requirements for interacting with external build services and file-modification utilities; and
(4) an \emph{Output and Verification Module} enforcing strict formatting constraints and external build-validation protocols.
All evaluated models are provided with identical prompts without model-specific tuning to ensure fair comparison across systems.

To ground our design choices in established research practice, we note that prior work on automated program repair generally operates at two dominant granularities: 
(1) patch-level editing and 
(2) full file or regeneration. 
These two approaches represent the most established and representative repair paradigms in LLM-based software engineering research~\cite{DBLP:conf/iclr/JimenezYWYPPN24, yang2025survey, zhang2026systematicliteraturereviewlarge}. 
Specifically, regenerating complete file contents emphasizes contextual completeness and global dependency consistency within modified artifacts~\cite{xu2025aligningobjectivellmbasedprogram}, whereas patch-based repair prioritizes localized edit precision and operational efficiency, following the canonical diff-style format widely adopted in automated repair benchmarks~\cite{yang2025survey, DBLP:conf/iclr/JimenezYWYPPN24}. 

Although intermediate granularities (\eg function-level or block-level generation) could potentially provide hybrid trade-offs, we deliberately focus on these two boundary-case paradigms to provide clear and interpretable insights into model sensitivity with respect to context preservation versus generation precision.  
Building on these established paradigms, we design two prompt configurations corresponding to the experimental repair strategies to guide LLM-driven repair and evaluate how different editing granularities influence repair performance.

(1) \textbf{Full File Generation Prompt.}
This prompt instructs the LLM to regenerate the complete files that are inferred to be related to the current build failure.
The model outputs the entire revised file, using a header-style notation such as ``\texttt{===FILE===: src/module/config.c}'' followed by the complete post-repair content and a closing ``\texttt{End of file}'' marker.
The prompt explicitly emphasizes three guiding principles.
\begin{itemize}
    \item \textit{Minimalism:} Modify only what is necessary to repair the failure.
    \item \textit{Completeness:} Always output the entire file to ensure syntactic correctness and reproducibility.
    \item \textit{Style preservation:} Retain the original code structure and comment layout.
\end{itemize}

(2) \textbf{Patch Generation Prompt.}
In contrast, the patch-based prompt directs the model to produce line-level modifications following the unified-diff convention used by Git.
The prompt explicitly enforces the use of valid file headers and hunk specifications to ensure the generated patches can be automatically applied through standard diff utilities by \name{}.
Each patch begins with a file-level header such as
``\texttt{diff --git a/<relpath> b/<relpath>}'' and one or more hunk headers of the
form ``\texttt{@@ -<start>[,<len>] +<start>[,<len>] @@}'', followed by line-level
edits where lines prefixed with ``\texttt{-}'' denote deletions, ``\texttt{+}'' additions,
and ``\texttt{ }'' contextual lines.
This format enables the \textit{Content Modification Tool} to apply the model's edits precisely to the target files without ambiguity.
To ensure robustness, \name{} includes a lightweight validation step within this tool to check the structural integrity of generated patches and automatically correct simple formatting inconsistencies (\eg malformed line prefixes or headers) before application.




Regardless of the repair strategy, the Context Module is programmatically reconstructed at each iteration using the execution-feedback components described in Section~\ref{subsubsec: iterative reasoning}.
This design preserves reasoning continuity and enables the model to reflect on prior decisions while adapting its repair hypothesis across iterations, thereby reducing redundant edits.
The complete prompt templates for both configurations, together with representative interaction traces, are provided in Appendix~\ref{append1} to facilitate reproducibility.
To assess the robustness of the prompt formulation, we further conduct a prompt-sensitivity study, where we systematically ablate individual prompt modules and evaluate performance variations.
Details are presented in Section~\ref{subsec: rq5}.

\section{Experiments}

We conduct an extensive evaluation across six representative large language models (LLMs) as the primary study models, covering both commercial API-based systems (GPT-5, GPT-5-mini, GPT-4o, Claude Sonnet 4.5, Qwen3-max) and open-source counterparts (DeepSeek V3), to systematically assess the effectiveness of \name{} in the context of cross-ISA build repair.
The experiments aim to answer the following research questions:
\begin{itemize}
    \item RQ1: How do current large language models perform in repairing cross-ISA build failures?
    \item RQ2: Does iterative feedback improve the repair performance of LLMs compared with single-shot evaluation?
    \item RQ3: How do \textit{Full File Generation} and \textit{Patch Generation} repair strategies affect the overall repair outcomes?
    \item RQ4: How sensitive is \name{} to variations in prompt design and module configurations?
    \item RQ5: How does an LLM complete the end-to-end repair and build process within \name{}?
\end{itemize}

To ensure a fair comparison, all models are evaluated under identical settings, including the same package corpus, prompt templates, tool interfaces, and iteration limits. 
Each model interacts with \name{} through a unified client interface that follows the Model Context Protocol (MCP), enabling consistent tool invocation and logging across runs.
We set the per-iteration tool-invocation limit at $T_{max}=20$ and the maximum repair iterations at $N_{max}=3$, which represent an empirically grounded trade-off between repair success and computational cost (detailed in Section~\ref{subsec:nmax_tmax_analysis}).

Regarding inference settings, since temperature control is not uniformly exposed across all commercial and open-source systems, we maintain the default configurations for each model to avoid biasing model-specific sampling. 
Repair outcomes are validated through external builds in OBS, which provide a consistent execution environment for all models. 
Each model is evaluated once per package under a fixed configuration, and the reported results reflect performance under these standardized settings.

\subsection{Task Formulation}

Given a software package $P$ that successfully builds on a source architecture $A_s$ (\eg x86\_64) but fails on a target architecture $A_t$ (\eg aarch64), the objective is to automatically repair $P$ so that it can be successfully rebuilt and verified on $A_t$.
Formally, let $\mathcal{B}(P, A) \rightarrow \{\texttt{succeed}, \texttt{failed}\}$ denote the build outcome of package $P$ under architecture $A$.
The goal is to find or evaluate a repair function $f_\theta$ satisfying:

\[
\mathcal{B}(f_\theta(P, A_s, A_t), A_t) = \texttt{success},
\]

where $f_\theta$ represents a large language model (LLM) that performs reasoning and modification across code, configuration, and build metadata.

Each task instance consists of the complete source package (including source archive, specification and metadata, and build scripts) and its corresponding architecture-specific build log.
The model is required to analyze build failures, infer root causes, and generate modifications that enable successful build on the target architecture.
This process may involve multiple repair iterations, where the model revises its previous repair content based on feedback from the latest build results.
The iterative process terminates when the package is successfully built or the maximum number of iterations is reached.

\subsection{Evaluation Metrics}
We evaluate both the \textbf{effectiveness} and \textbf{efficiency} of model-driven repair. 
The following metrics are adopted in \name{}:

\begin{itemize}
    \item \textbf{Build Success Rate}: the percentage of packages that are successfully built on the target architecture within $N_{\max}$ iterations.
    \item \textbf{Average Repair Time (min)}: the average time a package takes until successful build or termination.
    \item \textbf{Average Token Consumption (K)}: the average number of input and output tokens the model consumes for each package during the entire repair process.
\end{itemize}

This formulation provides a clear and measurable framework for assessing whether LLMs can understand, adapt, and repair software packages in cross-ISA migration scenarios, emphasizing their reasoning depth, contextual utilization, and cost-effectiveness.

\subsection{RQ1: Overall Performances on \name}

\begin{table*}[t]
  \centering
  \caption{Performance of LLMs on cross-ISA build failures in both migration directions.
    For each model, \emph{Success} indicates the number of packages that are successfully built;
    \emph{Success Rate} corresponds to Build Success Rate;
    \emph{Avg Time (min)} corresponds to Average Repair Time; 
    \emph{Avg Tokens (K)} corresponds to Average Token Consumption.}
  \label{tab:cross-isa-performance}
  \setlength{\tabcolsep}{4pt}
  \renewcommand{\arraystretch}{1.15}
  \begin{tabular*}{\textwidth}{@{\extracolsep{\fill}} l c l c c c c @{}}
    \toprule
    \textbf{Direction} & \textbf{Total} & \textbf{Model} & \textbf{Success} & \textbf{Success Rate (\%)} & \textbf{Avg Time (min)} & \textbf{Avg Tokens (K)} \\
    \midrule
    \multirow{7}{*}{\textbf{\makecell[l]{x86\_64\\$\rightarrow$aarch64}}}
    & \multirow{7}{*}{163}
    & GPT-5           & \textbf{103} & \textbf{63.19} & 31.18 & 1830.91 \\
    && GPT-5-mini      & 47 & 28.83 & 13.80 & 1683.95 \\
    && Qwen3-max & 28 & 17.18 & 35.69 & 505.39 \\
    && GPT-4o          & 22 & 13.50 & 5.93 & 541.66 \\
    && Claude Sonnet 4.5 & 16 & 9.82 & 6.27 & 328.76 \\
    && DeepSeek V3     & 13 & 7.98 & 11.37 & 235.53 \\
    && Qwen2.5-3B-Instruct & 8 & 4.91 & 27.65 & 591.90\\
    \midrule
    \multirow{7}{*}{\textbf{\makecell[l]{aarch64\\$\rightarrow$x86\_64}}}
    & \multirow{7}{*}{105}
    & GPT-5           & \textbf{31} & \textbf{29.52} & 18.55 & 1518.66 \\
    && GPT-5-mini      & 28 & 26.67 & 14.37 & 1894.60 \\
    && Qwen3-max & 6  & 5.71 & 27.46 & 359.16 \\
    && GPT-4o          & 13 & 12.38 & 5.82 & 614.12 \\
    && Claude Sonnet 4.5 & 6  & 5.71 & 4.52 & 332.99 \\
    && DeepSeek V3     & 4 & 3.81 & 19.27 & 445.03 \\
    && Qwen2.5-3B-Instruct & 2 & 1.90 & 24.33 & 373.81\\
    \bottomrule
  \end{tabular*}
\end{table*}


To provide a more comprehensive evaluation, we further include a lightweight open-source small language model (SLM), Qwen2.5-3B-Instruct, as an additional baseline for RQ1.
Table~\ref{tab:cross-isa-performance} summarizes the overall repair results of these models across both migration directions.

\subsubsection{Cross-ISA Repair Accuracy}
Among all evaluated models, GPT-5 achieves the highest overall success rate in both migration directions, successfully builds 103 out of 163 failed packages \textbf{(63.19\%)} in the x86\_64$\rightarrow$aarch64 direction and 31 out of 105 packages \textbf{(29.52\%)} in the reverse aarch64$\rightarrow$x86\_64 migration.
Among other models, GPT-5-mini (28.83\%) also exhibits strong repair capabilities in the forward direction.
For the reverse direction (aarch64$\rightarrow$x86\_64), GPT-5-mini (26.67\%) and GPT-4o (12.38\%) remain competitive, showing moderate consistency across architectures.

These results indicate that while current LLMs demonstrate a promising ability to understand and repair cross-ISA build failures, their overall performance still leaves substantial room for improvement. 
We further analyze the failed repair cases and find that most failures can be attributed to three primary causes.
(1) \textbf{Limited comprehension of long or interleaved build logs.}
This often leads to the generation of redundant or cyclic tool calls, which eventually force termination upon reaching the predefined tool-call limit ($T_{max}$) in \name{}.
(2) \textbf{Incomplete output or premature truncation.}
The resulting repair solutions, though syntactically valid, lack functional completeness and thus fail to resolve build issues.
(3) \textbf{Incorrect tool invocation sequences.}
For example, after decompressing and modifying the source achieve, the LLM occasionally attempts to upload all files directly to the OBS without first recompressing the modified code.
This behavior triggers an error, yet the model fails to produce a corrective follow-up action and instead terminates the repair process prematurely.
A detailed analysis of tool invocation behaviors across LLMs is presented in Section~\ref{subsubsec: analysis of tool calls}.

Moreover, most models achieve higher success rates when migrating from x86\_64 to aarch64 than in the reverse direction.
This asymmetry may stem from the fact that recent industrial and research efforts predominantly focus on the forward migration path, allowing LLMs to accumulate more exposure and implicit knowledge related to this scenario.
Conversely, the reverse migration direction remains less explored, revealing limited understanding and adaptability.
The observed difference also mirrors the practical asymmetry in toolchain maturity and dependency availability between the two architectures.

\subsubsection{Efficiency and Token Utilization}
\label{subsubsec: efficiency and token}
In addition to repair accuracy, we further analyze the efficiency of model-driven repair in terms of average time and token consumption per package.
Across both migration directions, the average repair time ranges from approximately 5 to 36 minutes per package, reflecting notable variation in reasoning strategies and convergence stability.

GPT-5 achieves the highest success rate while maintaining reasonable efficiency, suggesting well-structured reasoning with moderate computational overhead.
GPT-5-mini generally exhibited moderate repair times (approximately 14 minutes per package) but consumed a higher number of tokens.
This pattern suggests that the model engages in extensive iterative reasoning and tool interactions, which improve robustness but incur higher computational overhead.
Despite its moderate token consumption, Qwen3-max still exhibits a relatively long average repair time of approximately 36 minutes per package.
GPT-4o and Claude Sonnet 4.5 complete repairs within 7 minutes, while their success rates remain limited because they often terminate early without fully resolving dependency or configuration inconsistencies.
In contrast, DeepSeek V3 demonstrates concise reasoning but lower success, indicating efficient yet less exhaustive exploration.
Our analysis shows that most of this time is spent on repeatedly invoking the \textit{Build Result Check} tool to verify intermediate build outcomes.
This excessive verification overhead substantially prolongs the repair cycle without contributing to additional successful cases.

Overall, the results reveal a trade-off between reasoning depth and efficiency.
Models that maintain longer reasoning chains and richer tool interactions tend to achieve higher success rates but consume more time and tokens.
Conversely, faster models often exhibit insufficient context retention or under-exploration of repair strategies.

\subsubsection{Feasibility under Resource Constraints}
\label{subsubsec: feasibility for scenarios}

To further evaluate the applicability of \name{} in resource-constrained settings, we include an open-source small language model (SLM), Qwen2.5-3B-Instruct, as a lightweight baseline. 
This model can be deployed locally, eliminating API costs and providing stronger control over data privacy. 
As shown in Table~\ref{tab:cross-isa-performance}, Qwen2.5-3B-Instruct achieves success rates of 4.91\% for x86\_64$\rightarrow$aarch64 and 1.90\% for aarch64$\rightarrow$x86\_64. 
Although its repair capability is substantially lower than that of frontier-scale models, it is still able to complete a small number of end-to-end repair tasks within \name{}. 
This result suggests that the standardized tool interfaces and iterative feedback loop in \name{} also enable smaller models to participate in complex multi-stage build repair workflows.
At the same time, the relatively long repair time of the SLM indicates that lower-capacity models may require more trial-and-error before converging, even when their final success rate remains limited. 
Overall, these findings show that \name{} can support both frontier LLMs and locally deployable SLMs, providing a practical option for developers operating under computational, privacy, or financial constraints.

\subsection{RQ2: Effect of Iterative Feedback}

\begin{table*}[t]
\centering
\caption{Comparison of Bare-LLM single-shot baseline vs. \name{} (agentic) performance. 
Bare-LLM corresponds to a single-shot repair attempt with minimal tool usage. 
}
\label{tab:bare_llm_comparison}
\setlength{\tabcolsep}{4pt}
\renewcommand{\arraystretch}{1.15}
\begin{tabular*}{\textwidth}{@{\extracolsep{\fill}} l c l c c c c @{}}
\toprule
\textbf{Direction} & \textbf{Total} & \textbf{Model} & \textbf{Success} & \textbf{Success Rate (\%)} & \textbf{Avg Time (min)} & \textbf{Avg Tokens (K)} \\
\midrule

\multirow{6}{*}{\textbf{\makecell[l]{x86\_64\\$\rightarrow$aarch64}}} & \multirow{6}{*}{163}
& GPT-5 (Bare) & 10 & 6.13 & 2.13 & 17.35 \\
& & GPT-5-mini (Bare) & 7 & 4.29 & 1.78 & 15.21 \\
& & Qwen3-max (Bare) & 3 & 1.84 & 2.41 & 12.49 \\
& & GPT-4o (Bare) & 1 & 0.95 & 1.40 & 10.81 \\
& & Claude Sonnet 4.5 (Bare) & 9 & 5.52 & 1.51 & 12.79 \\
& & DeepSeek V3 (Bare) & 2 & 1.23 & 2.00 & 11.18 \\
\midrule
\multirow{6}{*}{\textbf{\makecell[l]{aarch64\\$\rightarrow$x86\_64}}} & \multirow{6}{*}{105}
& GPT-5 (Bare) & 6 & 0.95 & 1.13 & 12.57 \\
& & GPT-5-mini (Bare) & 3 & 2.86 & 1.98 & 12.37 \\
& & Qwen3-max (Bare) & 2 & 1.90 & 2.05 & 13.01 \\
& & GPT-4o (Bare) & 2 & 1.90 & 1.62 & 11.04 \\
& & Claude Sonnet 4.5 (Bare) & 1 & 0.95 & 1.03 & 12.75 \\
& & DeepSeek V3 (Bare) & 1 & 0.95 & 1.91 & 12.33 \\
\bottomrule
\end{tabular*}
\end{table*}

To contextualize the performance of \name{} and isolate the contribution of agentic iterative reasoning, we first establish a \textit{Bare LLM} baseline. 
In this single-shot setting, models are provided with the same initial build logs and package metadata but are restricted from using any external tools or receiving iterative feedback from the Build Service.
As shown in Table~\ref{tab:bare_llm_comparison}, the performance gap is substantial. 
Without the aid of agentic orchestration, GPT-5 achieves a success rate of only 6.13\% in the forward migration direction. 
In contrast, when operating within the \name{} framework, its success rate increases to 63.19\%, representing a 10.3$\times$ improvement. 
Similar gaps are observed across all evaluated models and both migration directions.
This significant disparity underscores that cross-ISA build repair is not a simple code-fixing task that can be resolved through pre-trained knowledge alone. 
Instead, successful repair requires dynamic tool orchestration and verifiable feedback loop to localize remaining failures, revise repair hypotheses, and maintain multi-file consistency. 
The following analysis further details how models leverage this iterative feedback to progressively converge toward successful repairs.

Table~\ref{tab:iterative-feedback} summarizes the effect of iterative feedback on cumulative repair success across three iterations in \name{}.
Each iteration reuses the latest build log and prior repair output as contextual feedback, allowing the model to refine its reasoning and avoid repeating ineffective edits.

Across both migration directions, iterative feedback consistently improves repair outcomes, confirming that LLMs benefit from exposure to updated diagnostic information.
In the x86\_64$\rightarrow$aarch64 direction, GPT-5 shows the most pronounced improvement, rising from 36.81\% after the first iteration to 63.19\% after the third (26.38 points).
GPT-5-mini and Qwen3-max also achieve steady gains of 13.49\% and 11.04\%, respectively, while DeepSeek V3 and GPT-4o improve more modestly.
By contrast, Claude Sonnet 4.5 shows no change across iterations, suggesting limited feedback utilization.
Our analysis reveals that Claude Sonnet 4.5 typically modifies file contents only during the first iteration.
In later iterations, the model merely inspects the files and directly re-uploads the package to the build service without making any further edits.
This behavioral pattern suggests that Claude Sonnet 4.5 fails to reinterpret the updated auxiliary context (new build logs and prior patch results) and instead repeats its initial reasoning trajectory.

\begin{table*}[t]
  \centering
  \caption{Iteration-wise improvement in build success rate on \name{}.
  Iter-1, Iter-2, and Iter-3 denote the cumulative build success rates after the first, second, and third iterations, respectively.
  $\Delta$(3–1) represents the improvement between the first and third iterations.}
  \label{tab:iterative-feedback}
  \setlength{\tabcolsep}{6pt}
  \renewcommand{\arraystretch}{1.15}
  \begin{tabular*}{\textwidth}{@{\extracolsep{\fill}} l c l c c c c @{}}
    \toprule
    \textbf{Direction} & \textbf{Total} & \textbf{Model} & \textbf{Iter-1 (\%)} & \textbf{Iter-2 (\%)} & \textbf{Iter-3 (\%)} & \textbf{$\Delta$(3–1)} \\
    \midrule
    \multirow{6}{*}{\textbf{\makecell[l]{x86\_64\\$\rightarrow$aarch64}}}
    & \multirow{6}{*}{163}
    & GPT-5           & 36.81 & 48.47 & 63.19 & $\uparrow$\textbf{26.38} \\
    && GPT-5-mini      & 15.34 & 20.25 & 28.83 & $\uparrow$13.49 \\
    && Qwen3-max & 6.13 & 9.82 & 17.18 & $\uparrow$11.04 \\
    && GPT-4o          & 4.91 & 12.88 & 13.50 & $\uparrow$8.59 \\
    && Claude Sonnet 4.5 & 9.82 & 9.82 & 9.82 & $\uparrow$0 \\
    && DeepSeek V3     & 3.68 & 6.13 & 7.98 & $\uparrow$4.29 \\
    \midrule
    \multirow{6}{*}{\textbf{\makecell[l]{aarch64\\$\rightarrow$x86\_64}}}
    & \multirow{6}{*}{105}
    & GPT-5           & 11.43 & 16.19 & 29.52 & $\uparrow$\textbf{18.10}\\
    && GPT-5-mini      & 12.38 & 18.10 & 26.67 & $\uparrow$14.29 \\
    && Qwen3-max & 0.95 & 0.95 & 5.71 & $\uparrow$4.76 \\
    && GPT-4o          & 0.95 & 8.57  & 12.38 & $\uparrow$11.43 \\
    && Claude Sonnet 4.5 & 4.76 & 4.76 & 5.71  & $\uparrow$0.95 \\
    && DeepSeek V3     & 0.95 & 0.95 & 3.81 & $\uparrow$2.86 \\
    \bottomrule
  \end{tabular*}
\end{table*}

In the reverse direction (aarch64$\rightarrow$x86\_64), the overall trend remains consistent.
GPT-5 again demonstrates the largest improvement (18.10 points), followed by GPT-5-mini (14.29 points) and GPT-4o (11.43 points).
DeepSeek V3 and Qwen3-max show limited yet positive gains of 2.86\% and 4.76\%, indicating that iterative feedback remains helpful even when model reasoning capacity is restricted.

These observations demonstrate that iterative feedback significantly enhances model reliability and cross-iteration learning, particularly for models with stronger contextual reasoning and tool-usage consistency.

\subsubsection{How LLMs Succeed or Fail During Iterative Repair}
While iterative feedback generally improves repair outcomes, its effectiveness varies across models and software packages. 
Based on the observed behaviors, we categorize repair patterns into three types:

(1) Immediate convergence through explicit log signals (successful in a single iteration).
This category includes packages whose build logs provide clear and localized error evidence.
For instance, in the case of \texttt{abi-compliance-checker}~\footnote{\url{https://build.opensuse.org/package/show/openSUSE:Factory/abi-compliance-checker}}, the build log reports
``\texttt{ERROR: can't compile libsample\_c v.2: `libsample\_c/libsample.v2/build-log.txt'},''
which explicitly indicates a missing or incompatible C library.
The message also clarifies that the failure occurs in the testing script rather than in the build or packaging stages.
With this direct diagnostic signal, GPT-5 and GPT-5-mini successfully associate the error with the corresponding \texttt{Makefile} and configuration files, adjust the \texttt{gcc} compilation options (for example, by adding \texttt{-fPIC} and \texttt{-fpermissive}), and regenerate the package correctly within a single iteration.
When the log exposes a clear causal link between the failure point and its configuration source, the repair process becomes almost deterministic, and models with sufficient reasoning capability can complete the repair without further iterations.

(2) Gradual improvement through accumulated contextual reasoning (successful after multiple iterations).
The build failures of these packages are distributed across multiple components or phases, requiring the model to progressively reason across iterations and refine its repair hypotheses.
For example, the \texttt{python-ironic-inspector-client}~\footnote{\url{https://build.opensuse.org/package/show/openSUSE:Factory/python-ironic-inspector-client}} package initially fails because several Python 3.13 dependencies are unresolved in the cross-ISA environment.
The build log repeatedly reports missing modules such as \texttt{python3-oslo.i18n} and incomplete setup configurations.
In the second iteration, GPT-5 successfully utilizes the previous failure log as contextual feedback, automatically infer the missing runtime requirements, and insert them into the \texttt{.spec} file under the \texttt{BuildRequires} and \texttt{Requires} sections.
It also corrects subtle macro inconsistencies introduced by the target architecture, ensuring a consistent Python packaging environment.
After these two iterations, the package is successfully rebuilt on the x86\_64 platform.
This case demonstrates that iterative reasoning allows the model to progressively uncover hidden dependencies and refine its repair strategies when the failure causes span multiple build stages.

(3) Non-convergent or degenerate repair loops (failed after multiple iterations).
Some packages remain failed to build even after three iterations.
Beyond general comprehension limitations, these persistent failures are often caused by procedural deficiencies.
For example, the llm may correctly decompress and modify the source files but repeatedly invoke the upload tool without recompressing the modified code.
This leads to packaging errors, after which the model fails to produce further corrective actions and instead terminates the repair process.
Such behavior indicates that limited procedural memory and tool sequence reasoning prevent recovery once the model enters an invalid operational path.

In summary, the contrast between one-shot, multi-round, and non-convergent repair cases reveals that iterative feedback enhances local reasoning.

\subsubsection{Tool Invocation Behavior Across LLMs}
\label{subsubsec: analysis of tool calls}

\begin{figure}[t]
  \centering
  \includegraphics[width=\linewidth]{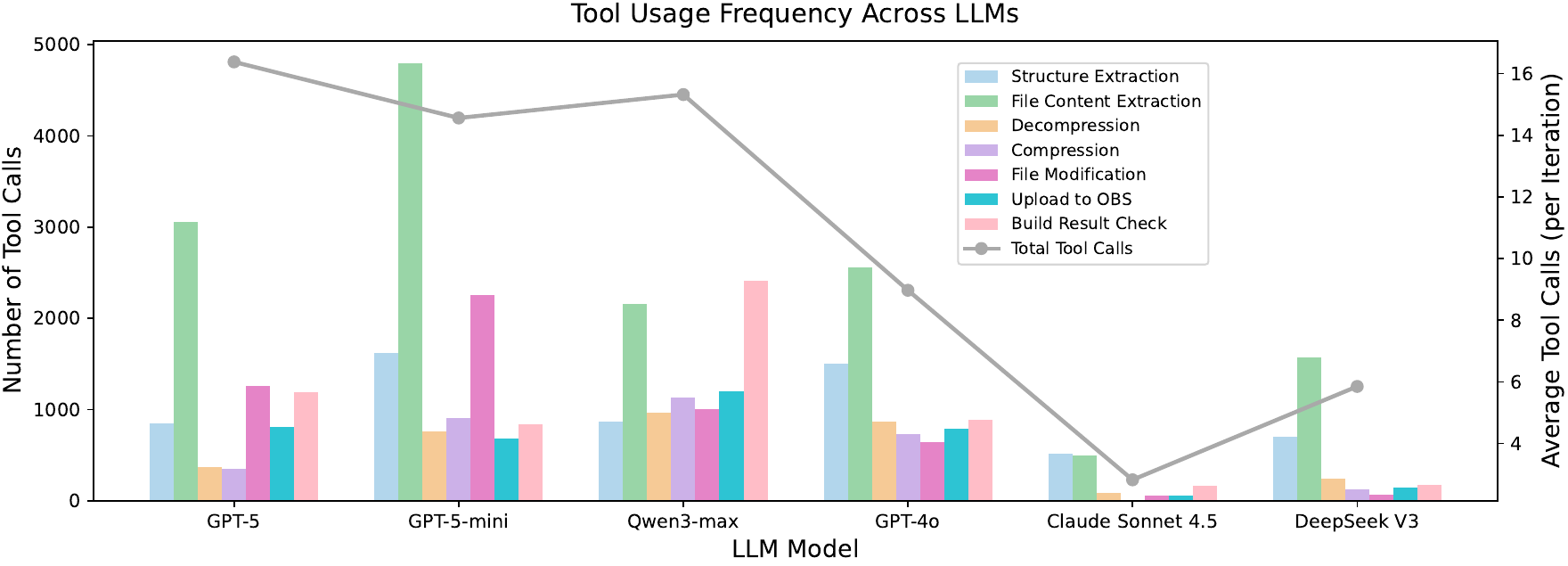}
  \caption{Comparison of tool invocation behavior across LLMs. The bars represent the total number of invocations for each tool per LLM, while the gray line indicates the average number of tool calls per iteration.}
    \label{fig: tool_usage}
\end{figure}

To further understand the behavioral characteristics of different models during cross-ISA repair, we analyze their tool invocation patterns as shown in Fig.~\ref{fig: tool_usage}. 
The bars indicate the total number of invocations for each tool across all 268 packages (including both migration directions), while the gray line represents the average number of tool calls per iteration, averaged over all repair attempts of each package.
To ensure stable execution and prevent infinite reasoning loops, the maximum number of tool invocations per iteration is capped at $T_{max}=20$. 
As further analyzed in Section~\ref{subsec:nmax_tmax_analysis}, this threshold provides a sufficient buffer for complex reasoning while effectively terminating non-convergent repair trajectories.

Overall, GPT-5 and GPT-5-mini exhibit the most active tool-invocation behaviors, adopting a more proactive reasoning strategy characterized by frequent verification and modification.
Among all tools, \textit{File Content Extraction} is invoked most often, indicating that these models frequently inspect specific source files to gather fine-grained, code-level evidence rather than relying solely on heuristic reasoning.
In addition, their frequent use of the \textit{File Modification} tool demonstrates that they can not only identify fault causes but also generate concrete repair content and automatically apply the modifications to the affected files.
Finally, both models consistently upload the modified packages to OBS for verification, suggesting that they possess a relatively strong capability for end-to-end task completion—integrating tool usage, contextual understanding, and procedural reasoning to accomplish complex repairs.

In contrast, GPT-4o, Claude Sonnet 4.5, and DeepSeek V3 show lower invocation frequencies, reflecting their limited understanding of the tool's functionality and a more conservative approach to iterative exploration.
Qwen3-max exhibits a distinctive behavior pattern, with an abnormally high frequency of \textit{Build Result Check} invocations.
Frequently revalidating the build state without engaging in substantive code-level reasoning results in redundant verification loops.

These findings reveal that LLMs differ not only in their linguistic reasoning capabilities but also in their operational strategies during tool-assisted repair, which is an important factor contributing to their divergent success rates in cross-ISA build repair.

\subsubsection{Justification of Iteration and Tool-Call Limits}
\label{subsec:nmax_tmax_analysis}
The iterative repair loop in \name{} is governed by two global hyperparameters: 
the maximum number of repair iterations ($N_{\max}=3$) and the maximum number of tool invocations ($T_{\max}=20$). 
These parameters are selected to balance repair effectiveness, computational cost, 
and runtime stability in large-scale autonomous software package repair.

\begin{figure}[t]
  \centering
  \includegraphics[width=\linewidth]{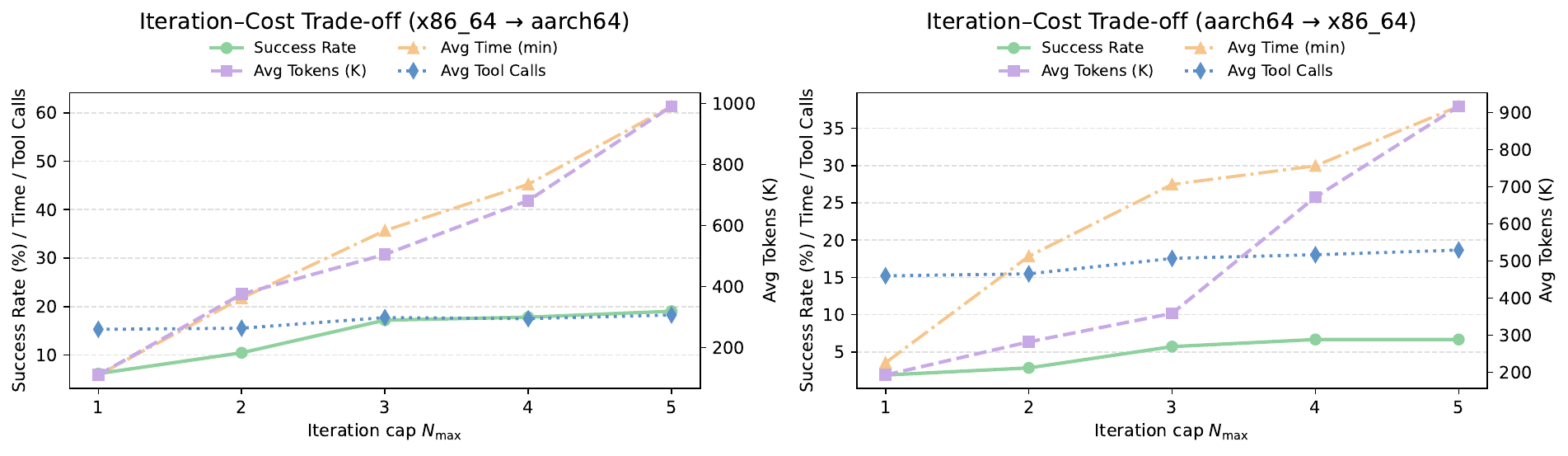}
  \caption{
  Iteration–cost trade-off under different iteration limit ($N_{\max}$) on \name{}.
  \emph{Success Rate} denotes the cumulative build success rate after each iteration cap.
  \emph{Avg Time (min)} and \emph{Avg Tokens (K)} correspond to average repair time and average token consumption per package, respectively.
  The \emph{Avg Tool Calls} indicate average number of tool invocations per iteration.
  }
  \label{fig:nmax_tradeoff}
\end{figure}

\textbf{Iteration limit ($N_{\max}$).}
To examine how repair success evolves across iterations and to assess the cost–benefit trade-off of additional attempts, we conduct an auxiliary analysis using Qwen3-max by varying $N_{\max}$ from 1 to 5. 
Figure~\ref{fig:nmax_tradeoff} illustrates the evolution of the \emph{Build Success Rate} and \emph{Avg Tokens (K)} as the iteration limit increases.
Across both migration directions, the majority of successful repairs occur within the first three iterations. 
In the x86\_64$\rightarrow$aarch64 migration, the cumulative success rate increases sharply from 6.13\% ($N_{\max}=1$) to 17.18\% ($N_{\max}=3$), capturing over 90.3\% of the successful cases found at $N_{\max}=5$. 
A similar trend is observed in the reverse direction, where 85.7\% of total successes are achieved within the first three iterations.

Extending the iteration budget to $N_{\max}=5$ yields only marginal gains while incurring a disproportionate escalation in computational cost.
For example, in the x86\_64$\rightarrow$aarch64 setting, the average token consumption per package increases from 505.39K at $N_{\max}=3$ to 991.24K at $N_{\max}=5$. 
The corresponding average repair time climbs from 35.69 minutes to 61.36 minutes.
A similar pattern is observed in the reverse direction, where the average repair time rises from 27.46 minutes at $N_{\max}=3$, and up to 37.94 minutes at $N_{\max}=5$, while token consumption also increases substantially.
This diminishing-return pattern indicates that most actionable diagnostic signals from build logs are exploited within the first few repair iterations, and further iterations primarily incur redundant verification or exploratory tool calls that significantly increase cost without proportionate gains.
Based on this empirical trade-off, we set $N_{\max}=3$ in the main experiments.

\textbf{Per-iteration tool-call limit ($T_{\max}$).}
During each iteration, the model may orchestrate diverse external tools, including file content extraction, modification, and build verification.
To prevent excessive or cyclic tool usage and maintain computational stability, we cap the number of tool invocations per iteration at $T_{\max}=20$. 
To empirically assess whether this threshold constrains model behavior, we analyze the average number of tool calls per repair attempt across different $N_{\max}$
using Qwen3-max.
As summarized in Figure~\ref{fig:nmax_tradeoff} and the accompanying statistics,
successful repairs typically require between 15 and 18 tool calls per iteration on average.
Specifically, for the x86\_64$\rightarrow$aarch64 direction, the average number of tool calls per iteration ranges from 15.27 to 18.25.
For the aarch64$\rightarrow$x86\_64 direction, the corresponding values range from
15.21 to 18.66.
Across both migration scenarios, the average tool invocation frequency remains consistently below the cap of 20, even as the iteration budget expands.

In summary, the selection of $N_{\max}=3$ and $T_{\max}=20$ represents an empirically grounded elbow point that optimizes the trade-off between repair effectiveness, efficiency, and computational stability, and they remain consistent across all evaluated models.

\subsection{RQ3: Impact of Repair Strategy}
\label{subsec: impact of repair strategy}

\begin{figure}[t]
  \centering
  \includegraphics[width=\linewidth]{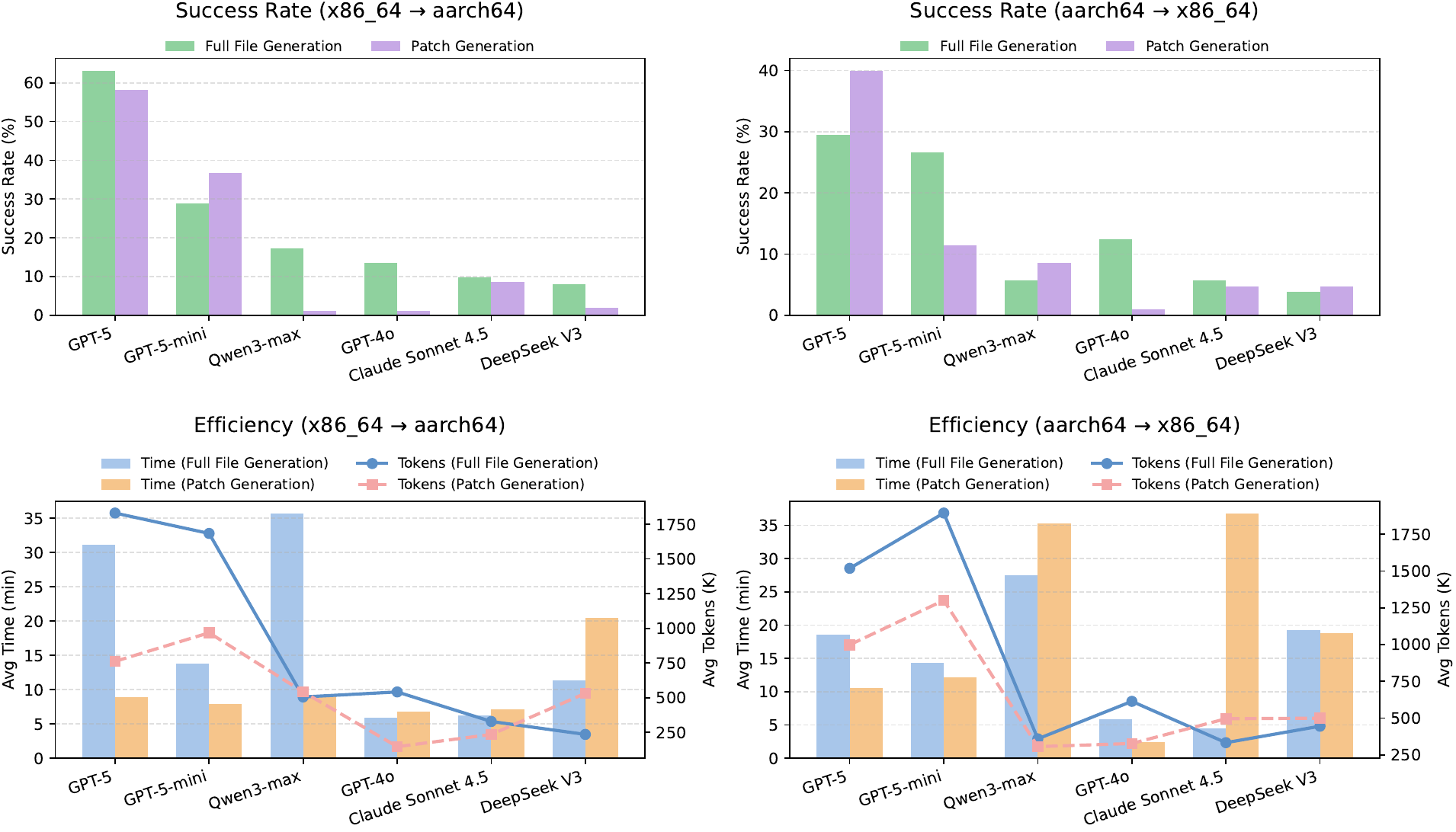}
  \caption{Comparison of two repair strategies (\textit{Full File Generation} vs. \textit{Patch Generation}) across six LLMs and two architecture migration directions.
  The upper row reports the \emph{Build Success Rate}, while the lower row presents \emph{Efficiency} in terms of \emph{Average Repair Time (min)} and \emph{Average Token Consumption (K)}.}
    \label{fig:repair strategy}
\end{figure}

To understand how different repair strategies affect the performance of automated build recovery, we compare two approaches adopted by \name{}: \textit{Full File Generation} and \textit{Patch Generation}.
In the \textit{Full File Generation} strategy, the large language model (LLM) regenerates the entire target file each time a modification is required, ensuring contextual completeness and dependency consistency at the cost of higher computational overhead. 
In contrast, the \textit{Patch Generation} strategy restricts edits to the modified segments.
The LLM outputs incremental diff-style changes (``+''/``-''), which are then automatically merged into the original file by \name{}.
This design substantially reduces long-sequence generation and mitigates potential truncation errors.
Fig.~\ref{fig:repair strategy} summarizes the comparative results across six large language models (LLMs) 
under two migration directions (x86\_64$\rightarrow$aarch64 and aarch64$\rightarrow$x86\_64). 

\begin{figure}[t]
  \centering
  \includegraphics[width=\linewidth]{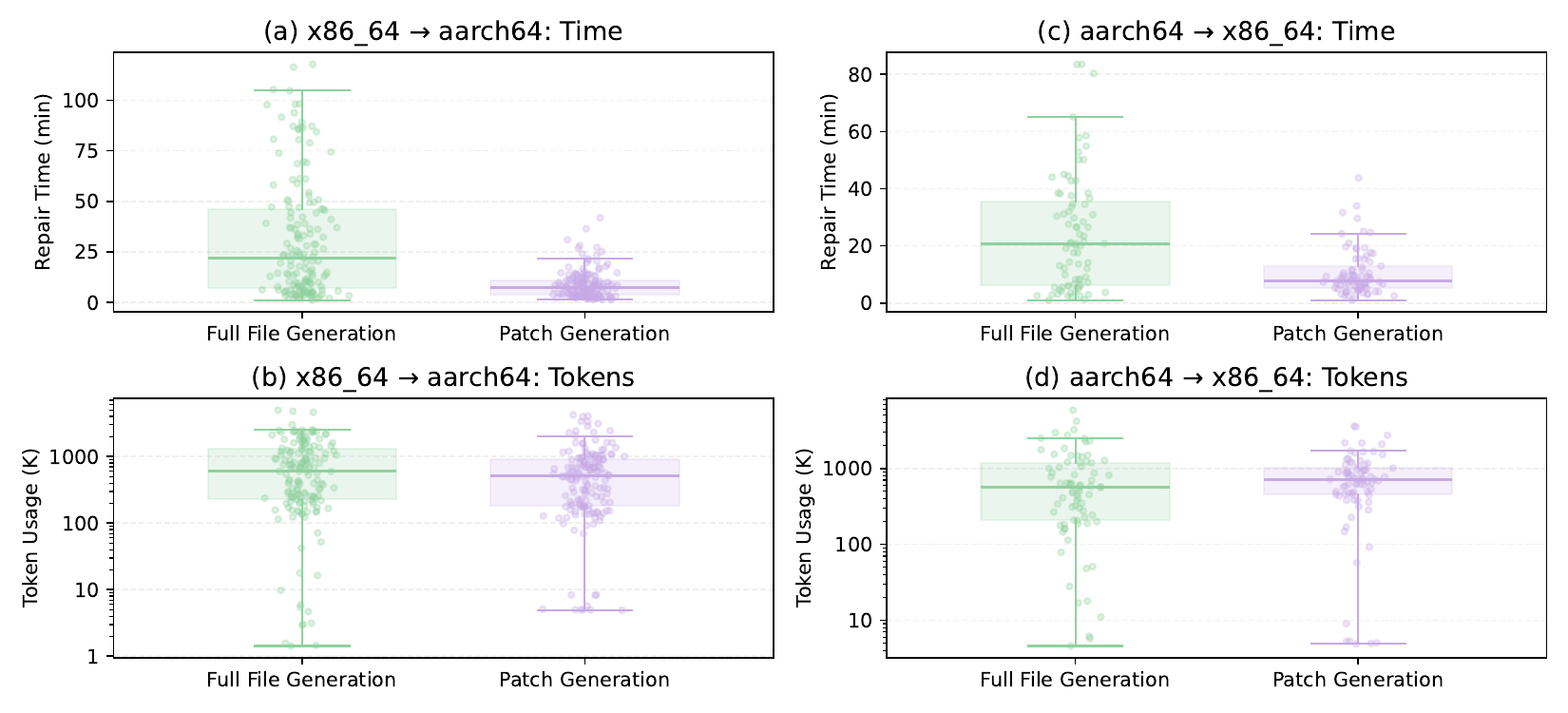}
  \caption{Per-package dispersion comparison between Full File Generation and Patch Generation for GPT-5 under two migration directions. The distributions are shown on log scale.}
  \label{fig:rq3-dispersion}
\end{figure}

Across both directions, the \textit{Patch Generation} strategy demonstrates significant gains in efficiency.
For instance, in the x86\_64$\rightarrow$aarch64 direction, GPT-5 reduces its average repair time from 31.18 to 8.93 minutes and token usage from 1830.91K to 761.88K.
Similarly, GPT-5-mini shortens the average repair time from 13.80 to 7.93 minutes and token usage from 1683.95K to 967.97K.
A comparable trend is observed in the reverse direction, which indicates that the \textit{Patch Generation} strategy achieves lower latency and a smaller token footprint across nearly all six models.
We further examine whether the observed efficiency improvements are broadly consistent across individual packages rather than driven by a small subset of extreme cases.
Fig.~\ref{fig:rq3-dispersion} presents per-package dispersion analysis for GPT-5 under both migration directions.
The boxplots show that \textit{Patch Generation} consistently shifts the entire distribution of repair time and token consumption downward.
The median and interquartile ranges are uniformly reduced, indicating that the efficiency gains are not limited to outliers, but hold for the majority of packages.

However, the success rate comparison shows a more nuanced pattern.
In the x86\_64$\rightarrow$aarch64 migration, almost all models achieve higher build success under the \textit{Full File Generation} strategy.
In the reverse aarch64$\rightarrow$x86\_64 direction, although GPT-5 shows a noticeable improvement under \textit{Patch Generation} (29.52\%$\rightarrow$40\%), most other models still perform slightly better with \textit{Full File Generation}.
This indicates that while the \textit{Patch Generation} strategy enhances efficiency, the \textit{Full File Generation} strategy generally remains more reliable in ensuring successful cross-ISA builds.

A plausible explanation for this discrepancy is twofold:
(1) The \textit{Patch Generation} strategy requires LLMs to output content that strictly conforms to patch formats (\eg diff structures and regular matching), and any format error may cause parsing failure in the toolchain.
Although \name{} incorporates an automated patch-validation module that verifies generated patches and corrects simple formatting inconsistencies prior to application, complex formatting deviations may still lead to patch application failures.
(2) \textit{Full File Generation} allows the LLM to regenerate the entire context, which can re-establish contextual consistency in scenarios with complex multi-file dependencies and is more conducive to successful cross-ISA builds.
Overall, these results reveal a fundamental trade-off between efficiency and completeness.
\textit{Patch Generation} substantially accelerates iterative repair and minimizes computational costs, whereas \textit{Full File Generation} offers stronger robustness and higher build success rate by reconstructing the full dependency context.

\subsubsection{Failure-Category Breakdown of Repair Success}
\label{subsec:rq3_failure_category}

\begin{table*}[t]
\centering
\caption{Repair success rates (\%) across failure categories for GPT-5 under the two repair strategies.
Results are reported separately for the two ISA migration directions.
}
\label{tab:rq3_category_breakdown}
\setlength{\tabcolsep}{4pt}
\renewcommand{\arraystretch}{1.15}
\begin{tabular*}{\textwidth}{@{\extracolsep{\fill}} l cc cc @{}}
\toprule
\multirow{2}{*}{\textbf{Failure Category}}
& \multicolumn{2}{c}{\textbf{x86\_64 $\rightarrow$ aarch64}}
& \multicolumn{2}{c}{\textbf{aarch64 $\rightarrow$ x86\_64}} \\
\cmidrule(lr){2-3} \cmidrule(lr){4-5}
& \textbf{Full} & \textbf{Patch} & \textbf{Full} & \textbf{Patch} \\
\midrule
Build Preparation Error           & 71.93 & \textbf{73.68} & \textbf{38.10} & 33.33 \\
Compilation Error                 & \textbf{62.69} & 49.25 & 18.37 & \textbf{32.65} \\
Packaging Error                   & 46.67 & \textbf{60.00} & 50.00 & \textbf{60.00} \\
Test Failure                      & \textbf{52.63} & 47.37 & 36.36 & \textbf{54.55} \\
Environment/Infrastructure Error  & \textbf{60.00} & 40.00 & 33.33 & 33.33 \\
\bottomrule
\end{tabular*}
\end{table*}

To examine whether different repair strategies exhibit sensitivity to the underlying causes of build failures, we perform a stratified analysis of repair success rates across the five primary failure categories defined in Table~\ref{tab:build-failure-classification}.

Table~\ref{tab:rq3_category_breakdown} presents the success rates of GPT-5 under the two repair strategies across both ISA migration directions. 
The results demonstrate that the relative advantage of a strategy is context-dependent. 
In the x86\_64 $\rightarrow$ aarch64 direction, \textit{Full File Generation} achieves higher success rates for categories such as \textit{Compilation Error} (62.69\%), \textit{Test Failure} (52.63\%), and \textit{Environment/Infrastructure Error} (60.00\%), while \textit{Patch Generation} remains slightly more reliable for \textit{Build Preparation Error} (73.68\%) and \textit{Packaging Error} (60.00\%).
Interestingly, in the aarch64 $\rightarrow$ x86\_64 migration, \textit{Patch Generation} significantly outperforms in \textit{Compilation Errors} (32.65\% vs. 18.37\%) and \textit{Test Failures} (54.55\% vs. 36.36\%).

These results indicate that the performance trade-off between the two strategies is not dominated by a single failure category.
Instead, different repair granularities exhibit varying strengths depending on the underlying nature of the build failure.
In particular, \textit{Patch Generation} tends to perform well when failures can be resolved through localized edits, while \textit{Full File Generation} may be more robust for cases that require broader contextual regeneration or structural adjustments.
For a more granular perspective, a detailed breakdown of success rates across the subcategories of build failures is provided in Appendix~\ref{append2}.

\subsection{RQ4: Prompt Sensitivity Analysis}
\label{subsec: rq5}
To examine whether the effectiveness of \name{} depends heavily on specific prompt formulations, we conduct a sensitivity analysis using Qwen3-max as the base model under the \textit{Full File Generation} strategy. 
The analysis consists of two parts: (1) automated prompt optimization and (2) prompt ablation experiments.

We first employ DSPy~\cite{khattab2023dspycompilingdeclarativelanguage}, a programmatic framework for prompt optimization, to automatically refine the repair instructions. 
Using a subset of 20 representative packages as a development set, we optimize the prompt with respect to end-to-end repair success. 
The resulting prompts exhibit only negligible textual differences from our manually designed version and yield no meaningful performance improvement. 
This observation suggests that the original prompt has already converged to a relatively stable formulation.

To quantify the impact of specific instruction components, we construct three variants of the original prompt:
\begin{itemize}
    \item \textbf{Variant A (Output Constraints):} We remove the strict requirement for full-file output and the associated formatting tags (\eg \texttt{===FILE:...===}).
    \item \textbf{Variant B (Operational Rules):} We ablate the \emph{Packaging \& Stop Rules}, including the requirement to re-compress the package root after decompression before uploading, and the enforced final ``compress$\rightarrow$upload'' tool-call sequence.
    \item \textbf{Variant C (Domain Knowledge):} We generalize the architecture descriptions, replacing specific mentions of ``x86\_64 to aarch64'' with generic terms like ``source to target architecture.''
\end{itemize}

\begin{table*}[t]
\centering
\caption{Prompt sensitivity analysis results using Qwen3-max (3 iterations). Variant A, B, and C correspond to the removal of output, operational, and domain-specific instructions, respectively.}
\label{tab:prompt_sensitivity}
\setlength{\tabcolsep}{5pt}
\renewcommand{\arraystretch}{1.15}
\begin{tabular*}{\textwidth}{@{\extracolsep{\fill}} l c c c c c @{}}
\toprule
\textbf{Direction} & \textbf{Total} & \textbf{Mode} & \textbf{Success Rate (\%)} & \textbf{Avg Time (min)} & \textbf{Avg Tokens (K)} \\
\midrule
\multirow{4}{*}{\textbf{\makecell[l]{x86\_64\\$\rightarrow$aarch64}}}
& \multirow{4}{*}{163}
& \name{}  & \textbf{17.18} & 35.69 & 505.39 \\
&& Variant A  & 16.56 & 29.03 & 400.89 \\
&& Variant B  & 11.04 & 32.95 & 545.44 \\
&& Variant C  & 15.34 & 28.12 & 435.25 \\
\midrule
\multirow{4}{*}{\textbf{\makecell[l]{aarch64\\$\rightarrow$x86\_64}}}
& \multirow{4}{*}{105}
& \name{} & \textbf{5.71} & 27.46 & 359.16 \\
&& Variant A & 4.76 & 27.43 & 471.98 \\
&& Variant B & 2.86 & 29.03 & 377.76 \\
&& Variant C & 5.71 & 19.17 & 451.93 \\
\bottomrule
\end{tabular*}
\end{table*}

As shown in Table~\ref{tab:prompt_sensitivity}, the overall repair performance remains relatively stable across most prompt variants. 
For x86\_64$\rightarrow$aarch64, removing output constraints (Variant A) or generalizing architecture descriptions (Variant C) leads to only moderate decreases in success rate, from 17.18\% to 16.56\% and 15.34\%, respectively. 
The largest degradation is observed in Variant B, where the success rate drops from 17.18\% to 11.04\%. 
A similar trend is observed for aarch64$\rightarrow$x86\_64, where Variant B again produces the lowest success rate (2.86\%).
Our error analysis suggests that the degradation in Variant B is primarily caused by violations of the packaging workflow rather than deficiencies in repair reasoning itself. 
In particular, for x86\_64$\rightarrow$aarch64, only 31 packages in Variant B invoked the re-compression step before uploading, compared with 91 packages under \name{}. 
The missing re-compression step frequently resulted in invalid uploads or stale artifacts, causing build failures even when the generated repair was otherwise plausible.

Overall, these results demonstrate that while specific engineering constraints in the prompt help ensure tool-chain compatibility, the core cross-ISA diagnostic and repair capabilities of the model are robust to reasonable variations in prompt formulation.

\subsection{RQ5: Case Study}

\begin{figure}[t]
  \centering
  \includegraphics[width=\linewidth]{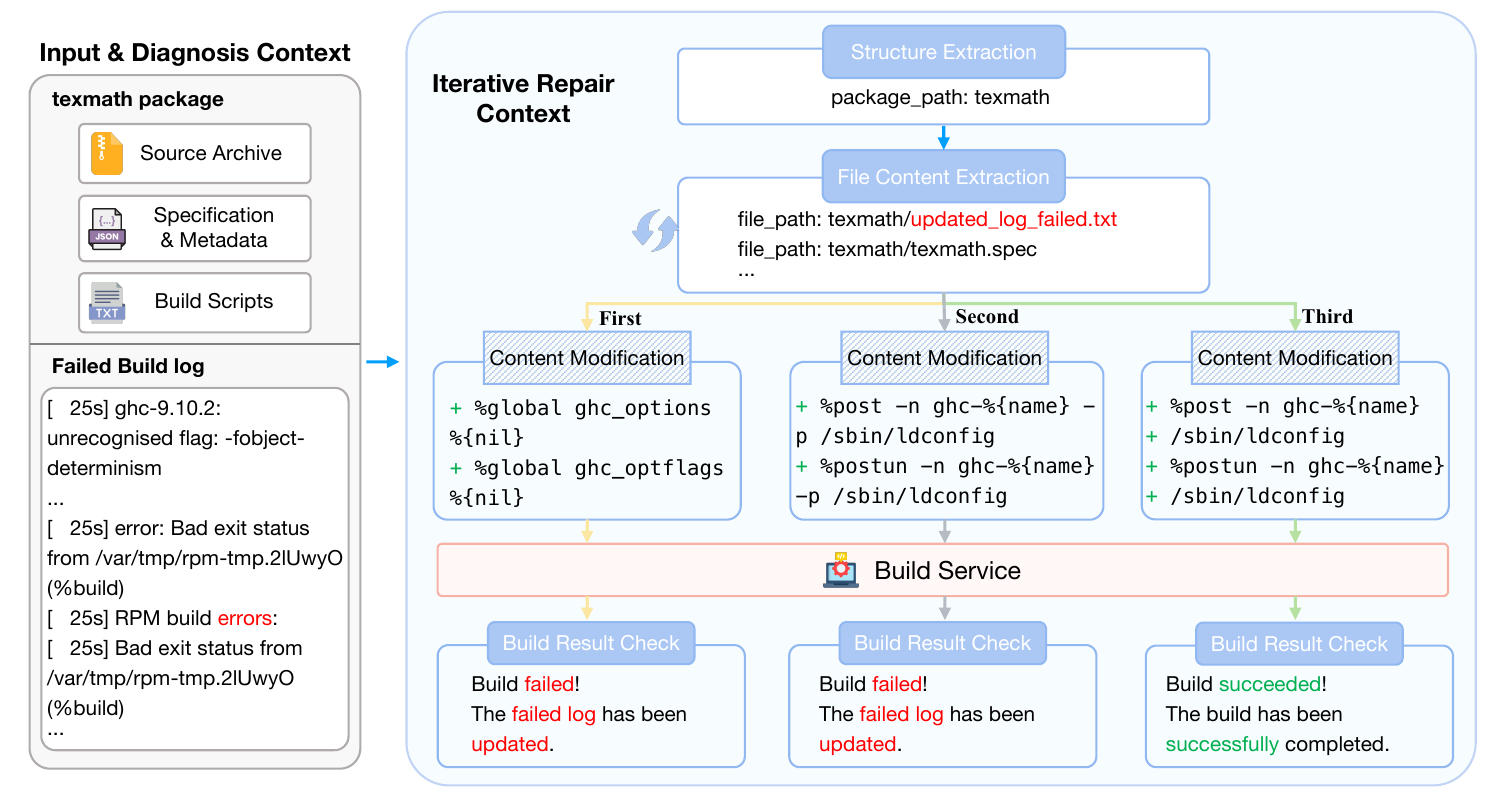}
  \caption{Iterative repair process of the \texttt{texmath} package migration under full-file generation.}
    \label{fig:case}
\end{figure}

As shown in Fig.~\ref{fig:case}, to illustrate how \name{} performs end-to-end iterative repair, we analyze the migration of the \texttt{texmath}~\footnote{\url{https://build.opensuse.org/package/show/openSUSE:Factory/texmath}} package from x86\_64 to aarch64.
We select \texttt{texmath} as an illustrative example because its failure is triggered by an architecture-specific compiler flag incompatibility, which is representative of the \textit{Compilation Error} category summarized in Table~\ref{tab:build-failure-classification}.
Moreover, the repair trajectory involves a complete and representative orchestration chain, including structure inspection, log retrieval, file modification, package upload, and build verification.
This self-contained case therefore provides a concrete example for visualizing how \name{} integrates diagnosis, repair generation, and external build feedback in an end-to-end iterative loop.

The initial build failure occurred during the \texttt{\%build} phase, where the GHC compiler aborted with an unsupported flag error:
\begin{verbatim}
ghc-9.10.2: unrecognised flag: -fobject-determinism
error: Bad exit status from /var/tmp/rpm-tmp.2lUwyO (%build)
\end{verbatim}
It indicates that the architecture-specific RPM macros (\texttt{ghc-rpm-macros}) injected a deterministic flag incompatible with the target compiler.
The case study demonstrates how GPT-5, starting from the original input (source archive, specification and metadata files, build scripts, and build log), autonomously identifies, applies, and verifies repair actions across three iterations, ultimately achieving a successful build.
We further compare the repair process under two editing granularities: \textit{Full File Generation} and \textit{Patch Generation}. 
Although both strategies are driven by the same failed build evidence and ultimately target the same incompatibility, they expose the model to different editing constraints.
As a result, they may follow different convergence trajectories and produce different concrete repairs, even when they remain consistent in root-cause understanding and final build outcome.

\subsubsection{Repair with Full File Generation}
We first describe the repair trajectory under full-file generation, where files identified as potential causes of failure are regenerated at each iteration, and the newly generated files directly replace the original files.

\textbf{First Iteration: global suppression of incompatible GHC flags.}
GPT-5 first invokes the \textit{Structure Extraction} tool to analyze the layout of the specification file (\texttt{texmath.spec}) and identify macro definitions near the build configuration.
It then applies the \textit{File Content Extraction} tool to read the complete failure log and confirm that the unsupported compiler flag \texttt{-fobject-determinism} was introduced through the macro \texttt{\%ghc\_optflags}.
Afterward, GPT-5 regenerates the entire specification file, overwriting the previous version.
Specifically, it redefines the macros \texttt{\%global ghc\_options} and \texttt{\%global ghc\_optflags} to \texttt{\%{nil}}, thereby disabling incompatible compiler flags inherited from the default \texttt{ghc-rpm-macros}.
Once the modifications are complete, GPT-5 invokes the \textit{Upload to Build Service} tool to upload the full package contents to the Open Build Service (OBS), followed by verification through the \textit{Build Result Check} tool.
The \textit{Build Result Check} tool returns the message: ``\textit{Build failed! The failed log has been updated.}''.
Subsequently, GPT-5 re-uploads the package to the OBS and then returns a response rather than invoking another tool.
It signals the end of the first iteration and confirms that the initial repair attempt is unsuccessful.

\textbf{Second Iteration: adding \texttt{ldconfig} scriptlets.}
In the second iteration, GPT-5 continues the repair process using the updated build log.
It first calls the \textit{Structure Extraction} tool to inspect the package layout, confirming the presence of the specification file, source archive (\texttt{texmath-0.13.tar.gz}), and the latest failed log.
Then, the \textit{File Content Extraction} tool is invoked to retrieve both the log and the specification file \texttt{texmath.spec}.
GPT-5 regenerates the specification file and appends post-install and uninstall scriptlets that refresh the runtime linker cache:
\texttt{\%post -n ghc-\%{name} -p /sbin/ldconfig} and \texttt{\%postun -n ghc-\%{name} -p /sbin/ldconfig}.
After uploading the modified package to OBS, the \textit{Build Result Check} tool again reports a failed build.
GPT-5 subsequently reads the updated build log, makes no additional modifications, and uploads the package once more.
The model then returns a non–tool-call response, ending the second iteration and indicating that further repair is required.

\textbf{Third Iteration: correcting scriptlet form.}
GPT-5 performs a final inspection of \texttt{texmath.spec} via \textit{File Content Extraction}, ensuring the previous additions remain intact.
It regenerates the specification file again, rewriting the scriptlets into body form to comply with RPM specification standards.
Specifically, it removes the \texttt{-p} shorthand and places the command \texttt{/sbin/ldconfig} on a new line following the header (\eg \texttt{\%post -n ghc-\%{name}}).
The modified package is uploaded again, and \textit{Build Result Check} returned: ``Build result: Build succeeded! The build has been successfully completed.''
GPT-5 then terminates with a non–tool-call response, marking the completion of the third iteration and the successful repair of the package.

\subsubsection{Repair with Patch Generation}
We next describe the repair trajectory under patch-based modification.
Instead of regenerating the entire file, GPT-5 produces Git-style unified diffs that update only the affected regions of the target file, which are then applied via the unified patch tool.

Although the overall repair workflow remains the same as in the full-file setting, the concrete editing trajectory may differ across granularities.
This is because a single build failure may admit multiple valid repairs, and LLM-based agents may select different yet functionally effective repair strategies depending on the editing space exposed to them.
In this example, the patch-based strategy converges through a localized modification within the \texttt{\%build} stage, rather than through successive whole-file revisions.

\textbf{First Iteration: filtering unsupported compiler flags via a wrapper.}
After retrieving the failed build log, GPT-5 identifies that the build aborts because the GHC compiler receives the unsupported flag \texttt{-fobject-determinism}.
Instead of modifying macro definitions globally, GPT-5 generates a localized patch that introduces a lightweight wrapper script to filter out the unsupported flag before invoking the system compiler.
The patch also adjusts the build configuration so that the compilation step uses the wrapper:

\begin{findingbox}
\begin{verbatim}
diff --git a/texmath.spec b/texmath.spec
--- a/texmath.spec
+++ b/texmath.spec
@@ -124,11 +124,19 @@
 %build
-%define cabal_configure_options -f+executable
-%ghc_lib_build
+# Create a wrapper to filter unsupported GHC flag
+cat > %{_builddir}/ghc-nodet << 'EOF'
+#!/bin/bash
+...
+exec /usr/bin/ghc "${args[@]}"
+EOF
+chmod +x %{_builddir}/ghc-nodet
+%define cabal_configure_options --with-compiler=%{_builddir}/ghc-nodet -f+executable
+%ghc_lib_build
\end{verbatim}
\end{findingbox}

The unified patch tool parses the diff headers and hunk ranges (\eg \texttt{@@ -l,s +l,s @@}) and applies the localized edits to \texttt{texmath.spec}.
After the patch is applied, the modified package is uploaded to OBS and the build status is checked.
The first verification returns a transient timeout due to the asynchronous build process, while the subsequent check confirms that the build has succeeded.

Overall, the patch-based repair follows the same closed-loop workflow as the full-file strategy, but expresses modifications as minimal unified diffs rather than regenerating entire files.
Importantly, although the patch-based and full-file trajectories produce different concrete edits, they remain consistent in diagnosis: both identify the incompatible flag \texttt{-fobject-determinism} as the key obstacle and aim to prevent it from affecting the effective GHC invocation during the target-architecture build.

\subsubsection{Summary and Lessons Learned}
This case highlights GPT-5's progressive reasoning ability in handling cross-ISA build failures through iterative refinement. 
Across the three \texttt{modify+upload} cycles, GPT-5 evolves from low-level syntactic correction to higher-level procedural understanding. 
The first iteration neutralizes macro-propagated compiler flags that cause architecture-specific incompatibilities; 
the second restores runtime environment consistency by introducing post-install and uninstall scriptlets; 
and the third refines these scriptlets into standard body form, ensuring compliance with RPM packaging conventions. 
This progression shows how GPT-5 incrementally assimilates build feedback and transforms diagnostic cues into precise configuration edits. 
By continuously grounding its reasoning in both specification files and failed build logs, GPT-5 autonomously closes the repair loop and converges toward a verified, reproducible build on the aarch64 architecture.

At the same time, this case also shows that different editing granularities can induce different repair trajectories for the same failure.
Under \textit{Full File Generation}, the model converges through three iterations of whole-file regeneration and progressive refinement, whereas under \textit{Patch Generation} it reaches build success with a single localized modification.
This difference should not be interpreted as inconsistent diagnosis; rather, it reflects that the same underlying failure may admit multiple valid repairs at different granularities.
Finally, the successful outcome in this case is determined by whether OBS reports a successful build.
Therefore, this case study demonstrates build-level repair effectiveness, rather than full semantic validation of all runtime functionalities of the repaired package.

\section{Related Work}
\subsection{Automated Build and Repair Systems}
Software build automation has long been a foundational topic in software engineering.
Traditional build systems such as \textit{Make}, \textit{CMake}, and \textit{Autotools} automate dependency resolution and compilation, yet they rely heavily on human intervention when failures occur.
Modern continuous integration (CI) infrastructures, such as openSUSE Build Service (OBS)~\footnote{\url{https://build.opensuse.org}}, Fedora Copr~\footnote{\url{https://copr.fedorainfracloud.org}}, and Debian's~\footnote{\url{https://buildd.debian.org/}} build farms, extend this process to large-scale package ecosystems, providing reproducible build environments and build logs for diagnostic purposes.
These platforms provide reproducible environments and diagnostic build logs, but they remain largely passive: they detect failures without performing automated repair.

\begin{table*}[t]
\centering
\begin{threeparttable}
\caption{Comparison of LLM-based orchestration and traditional paradigms across key capability dimensions for cross-ISA build repair. \CIRCLE, \LEFTcircle, and \Circle denote full support, partial or stage-specific support, and no support, respectively.}
\label{tab:orchestrator_motivation}
\setlength{\tabcolsep}{5pt}
\renewcommand{\arraystretch}{1.15}
\begin{tabular*}{\textwidth}{@{\extracolsep{\fill}}lccc@{}}
\toprule
\textbf{Capability Dimension} & \textbf{Heuristic-based} & \textbf{Static ML-based} & \textbf{\begin{tabular}[c]{@{}c@{}}LLM-based Orchestrator\\ (\name{})\end{tabular}} \\ \midrule

Heterogeneous-stage repair         
& \LEFTcircle \cite{DBLP:conf/icse/HassanW18,DBLP:conf/kbse/Zhang0H0Z22,DBLP:conf/issta/FanWW0SZ20} 
& \Circle
& \CIRCLE \\

Multi-file contextual reasoning   
& \Circle 
& \Circle 
& \CIRCLE \\

Dynamic tool orchestration        
& \Circle    
& \Circle    
& \CIRCLE \\

Architecture-aware adaptation     
& \Circle    
& \Circle    
& \CIRCLE \\

Iterative log-reflection          
& \Circle 
& \LEFTcircle \cite{DBLP:journals/tosem/NourryKSSK25}    
& \CIRCLE \\

End-to-end executable verification
& \Circle    
& \Circle    
& \CIRCLE \\

\bottomrule
\end{tabular*}

\begin{tablenotes}
\small
\item \textbf{\CIRCLE}: native support across the entire repair loop.  
\item \textbf{\LEFTcircle}: limited, rule-bound, or stage-specific support.  
\item \textbf{\Circle}: not supported.  
\end{tablenotes}
\end{threeparttable}
\end{table*}

Recent work has therefore explored automatic recovery, diagnosis, and repair of build failures~\cite{DBLP:conf/icse/Sun25,DBLP:conf/icse/HassanW18,DBLP:journals/infsof/MakC24,DBLP:conf/kbse/Zhang0H0Z22}.
Early approaches mainly relied on heuristic or static-analysis techniques to infer missing dependencies, misconfigured flags, recurring repair patterns, or narrowly scoped CI/build configuration issues.
For example, heuristic methods range from history-driven build-script repair to dependency-graph-based diagnosis of build errors, both of which provide only localized support for selected stages or failure manifestations~\cite{DBLP:conf/icse/HassanW18,DBLP:conf/issta/FanWW0SZ20,DBLP:conf/kbse/Zhang0H0Z22}.
Later studies increasingly incorporated data-driven or learning-based methods to predict build outcomes, classify failure causes, or recommend corrective actions from historical evidence~\cite{DBLP:conf/sigsoft/Zhang0C0Z19,DBLP:journals/tosem/NourryKSSK25}.
However, these methods remain narrow in scope: they typically target specific failure classes, individual configuration artifacts, or diagnosis from pre-collected logs, rather than end-to-end repair of executable software packages.

These limitations become more pronounced in cross-ISA settings.
Unlike conventional build failures that are often localized to a single stage or artifact, cross-ISA failures frequently propagate across environment setup, dependency resolution, compilation, and packaging, and involve interactions among heterogeneous artifacts such as configuration files, build scripts, source archives, and build logs.
As summarized in Table~\ref{tab:orchestrator_motivation}, prior heuristic-based and static ML-based paradigms provide at most partial support (\LEFTcircle) for selected aspects of this problem, such as stage-specific repair or build-log-based diagnosis, but generally lack unified support (\Circle) for multi-file contextual reasoning, dynamic tool orchestration, architecture-aware adaptation, and end-to-end executable validation.
Consequently, they are ill-suited to failures that must be diagnosed and repaired through continuous interaction with an executable target environment.

More recently, the emergence of large language models (LLMs) has revitalized the automation of build and repair processes.
Instead of executing a fixed procedural pipeline, agentic frameworks delegate repair decisions to the model itself, allowing it to dynamically select tools, plan repair steps, and evaluate results.
Representative systems in this paradigm include CXXCrafter~\cite{cxxcrafter}, RepairAgent~\cite{repairagent}, AutoCodeRover~\cite{zhang2024autocoder}, and VulDebugger~\cite{liu2025agentdebugs}.
These frameworks follow a tool-augmented design, where the LLM operates within a fixed \textit{perceive–think–act} loop (\ie observing the state, reasoning, selecting a tool, and applying it iteratively) while the outer control flow remains static.
Nevertheless, existing systems are primarily evaluated within single-architecture or repository-level contexts.
They lack systematic integration with external build services and seldom provide reproducible, end-to-end repair loops that include iterative verification or cross-environment validation.

Unlike these systems, \name{} integrates LLM-driven reasoning into a controlled and verifiable build environment.
It leverages the Model Context Protocol (MCP) to expose standardized tool interfaces (\eg structure extraction, file modification, build validation) that can be dynamically invoked by the model.
This design enables automated, end-to-end build repair at the package level, allowing reproducible experimentation across real-world build failures.
By combining structured tool orchestration with contextual iteration, \name{} advances automated build repair from isolated heuristics toward an agentic, fully verifiable workflow.

\subsection{Benchmarks and Evaluation Frameworks}
Recent benchmark suites assess the reasoning and editing capabilities of large language models on real software artifacts and enable reproducible comparisons.

Repository and issue level benchmarks establish test based evaluation with full repository context. 
SWE-bench~\cite{DBLP:conf/iclr/JimenezYWYPPN24} pairs real GitHub issues with corresponding pull requests and requires a model to modify the codebase so that tests pass, covering thousands of tasks from popular Python repositories. It emphasizes multi file reasoning inside realistic projects and uses executable tests as ground truth.
SWE-bench-Live~\cite{DBLP:journals/corr/abs-2505-23419} extends this setting with continuously updated instances curated from recent issues and provides per instance Docker images for reproducible execution. It targets contamination resistant, end to end evaluation under a live benchmark that evolves over time.
SWE-Gym~\cite{SWE-Gym} further supplies an interactive training and evaluation environment that packages tasks with pre installed dependencies and executable verification. 
It reports gains on SWE-bench variants by training agents and verifiers on agent trajectories collected in the environment.
Beyond issue resolution, Zhang et al.~\cite{zhang2025buildbench} introduces a benchmark that evaluates LLM-based agents on compiling real-world open-source software.

Program repair and fault localization benchmarks complement repository level evaluation with fine grained instances.
Defects4J~\cite{Defects4J} provides real bugs from Java projects within a controlled framework for reproducible testing, and has become a long-standing benchmark in software engineering research.
AgentFL~\cite{AgentFL} formulates fault localization as a multi agent process involving comprehension, navigation, and confirmation, while MemFL~\cite{yeo2025memfl} introduces an external memory that combines static project summaries and dynamic feedback collected across iterations to support multi round reasoning.

These benchmarks focus on iterative reasoning and fine grained repair analysis, but generally do not involve cross-ISA migration, build validation, packaging, iterative build–repair cycles, or cross platform compilation.
Within this landscape, \name{} targets cross architecture build repair. 
It evaluates whether LLMs can analyze build logs, reason over specification files and sources, and achieve a successful build under a controlled external service. 
In doing so, it complements existing benchmarks by shifting from static patch validation toward dynamic, system level reconstruction that involves configuration, dependency management, packaging, and verification on heterogeneous instruction set architectures.

\section{Discussion}

\subsection{Generality Across Architectures}
\label{generality across architectures}
Although \name{} currently focuses on migration between x86\_64 and aarch64, the methodology and pipeline are architecture agnostic.
These two architectures were selected primarily because they represent the most widely adopted and well maintained build ecosystems in mainstream Linux distributions, providing stable compiler toolchains and abundant package metadata for evaluation.
However, \name{} does not encode any architecture specific prior knowledge, nor does it rely on language model fine tuning tailored to particular instruction sets.
The framework evaluates a model’s ability to reason about build logs, configuration scripts, and dependency specifications, which are shared abstractions across all compilation targets.

Therefore, extending \name{} to other architectures, such as \texttt{riscv64}, \texttt{ppc64le}, or \texttt{s390x}, requires only substituting the package corpus and corresponding build environment on the same platform.
Since the pipeline is entirely automated and interacts with the build service through standardized APIs, the evaluation protocol remains unchanged.
This design allows \name{} to serve as a general benchmark for assessing LLM capabilities in cross architecture migration tasks, independent of specific hardware ecosystems.

\subsection{Potential Biases Introduced by OBS}
The Open Build Service (OBS) provides a reproducible and controlled environment for software package building and testing, but its centralized nature may introduce several experimental biases that warrant discussion.  

First, the reproducibility of builds on OBS depends on the stability of its repositories and mirrors.  
Minor variations in upstream dependencies, package versions, or project metadata can affect build outcomes and introduce temporal variability that may influence evaluation results.  
To mitigate this issue, all builds in our study are performed within an independent project to ensure consistent dependency resolution.  
In addition, to guarantee that all experiments share the same environment and dependency versions, we fix a specific timestamp and environment snapshot as the baseline for evaluation.  
The success of each repair is determined not only by the feedback from the OBS platform, but also the modifications recorded in the logs and the \textit{Build Result Check} tool.

Second, OBS enforces strict quotas and scheduling policies, which may influence runtime statistics such as measured repair time.  
Since different packages may be queued or executed on heterogeneous worker nodes, the reported build duration could reflect scheduling latency rather than actual model reasoning time.  
To address this, we measure repair time at the model level by calculating time differences solely from timestamps printed in the logs.  
This measurement only accounts for the duration between the first tool invocation and the last response within each iteration, excluding inter-iteration intervals, while treating external queuing delays as uncertain noise.  

In summary, OBS enables practical and large scale evaluation but also introduces potential variations in timing, dependency freshness, and system conventions.  
We recognize that considering these factors is essential for interpreting benchmark results and ensuring reproducibility when applying the framework to other architectures or build services.  
To minimize such potential sources of bias, \name{} adopts multiple corrective measures throughout the evaluation.  
Looking ahead, a promising extension is to maintain a self-hosted, containerized build environment using Docker or similar orchestration tools.  
Such an environment would provide stricter control over dependency versions, runtime isolation, and long-term reproducibility, further supporting the construction of a fully independent verification platform beyond public build services.

\subsection{LLM-Induced Randomness and Reproducibility}

Another source of uncertainty in \name{} arises from the inherent randomness of large language models (LLMs).  
Even when all external build conditions are held constant, model-level nondeterminism can cause variation in repair outcomes across independent runs.  
This randomness originates from several factors, including token sampling, context window truncation, and latent instability in multi-step reasoning.

To minimize stochastic behavior, all evaluated models are executed with deterministic inference settings, such as temperature fixed to zero and top-$p$ sampling disabled when possible.  
In addition, the model prompts and tool invocation schemas are strictly serialized to ensure that each iteration receives identical contextual inputs.  
However, nontrivial differences can still emerge due to internal randomness in decoding, variations in model updates from API providers, and the probabilistic nature of long-context attention.  
For instance, the same input log may lead the model to generate syntactically distinct yet semantically equivalent repairs, or conversely, omit critical modification lines that alter the build result.

To improve reproducibility, future iterations of \name{} can incorporate multiple randomized runs per package and report aggregate statistics such as confidence intervals of success rates.  
Moreover, logging model responses at every step allows exact replay of reasoning traces, enabling deterministic re-evaluation under fixed conditions.  
These extensions would facilitate more rigorous comparison between models and provide a foundation for studying robustness under stochastic inference.

\section{Threats to Validity}

Despite the controlled experimental setup of \name{}, several validity concerns remain that may affect the interpretation and reproducibility of our findings.  

\textit{Internal Validity.}
Internal validity concerns whether the observed repair outcomes faithfully reflect the reasoning and tool-use capabilities of the evaluated models.  
Since the build process involves multiple asynchronous components, a few uncontrolled factors (\eg transient network latency, temporary OBS queue congestion, or inconsistent log flushing) may lead to incomplete or misaligned feedback between iterations.  
We mitigate these risks by recording all tool invocations and system responses in structured logs, ensuring that every reasoning trace is fully recoverable.  
A potential implementation-related threat to internal validity arises from the strict formatting requirements of patch-based edits. 
\textit{Patch Generation} requires the LLM to output modifications that strictly comply with the unified-diff format; consequently, minor syntactic deviations may lead to parsing failures even if the underlying repair logic is semantically correct.
To mitigate this implementation-sensitive confound, \name{} incorporates an automated patch-validation module designed to recognize and repair common formatting inconsistencies, such as missing prefixes or malformed diff headers, prior to application. 
While this mechanism reduces failures caused by trivial formatting errors, more complex structural inconsistencies or contextual mismatches in the diff hunk may still result in unsuccessful patch applications.
Therefore, some failures under the \textit{Patch Generation} strategy may partially reflect the model's structured-output brittleness rather than a fundamental inability to reason about the build failure.

\textit{Build Validity.}  
Build validity relates to whether the evaluation metrics accurately reflect the constructs of interest.  
Our metrics focus on three measurable quantities: build success rate, mean repair time per package, and mean token consumption.  
Each metric is computed using locally captured timestamps and token logs to ensure consistency across different architectures and models.  
However, these measurements capture only the reasoning and repair phases rather than the full end-to-end runtime on OBS, which may slightly underestimate total operational latency.
Moreover, build success on OBS is used as the operational criterion for repair success in \name{}. 
While this criterion is practical, reproducible, and well aligned with the goal of package-build repair, it primarily validates buildability rather than full semantic correctness or runtime functional equivalence. 
A repaired package that successfully builds may still contain latent defects that are not exercised during the build process or may exhibit behavioral deviations at runtime. 
Therefore, our results should be interpreted as evidence of build-level repair effectiveness, rather than comprehensive validation of all downstream functionalities of the repaired package.


\textit{External Validity.} 
Although \name{} currently evaluates two widely deployed instruction set architectures (ISAs), namely x86\_64 and aarch64, the benchmark pipeline is designed to be ISA-agnostic at the interface level. 
In principle, extending the benchmark to another ISA does not require modifications to the prompt design or tool interfaces. 
However, we have not yet empirically validated repair performance on additional ISAs such as riscv64. 
Repair effectiveness may vary across ISAs due to differences in compiler ecosystems, packaging conventions, and—importantly—the richness of available build evidence. 
For emerging or less mature ISAs, historical build logs, community-maintained packaging macros, and toolchain stability may be comparatively limited, potentially reducing the quality of contextual signals available to the LLM.
In such cases, improved corpus curation or additional contextual resources (\eg historical repair examples or ISA-specific documentation) may further enhance repair performance. 
We leave a full-scale empirical evaluation on a broader range of ISAs as future work.
In addition, \name{} evaluates two established repair paradigms (Full File Generation, Patch Generation) , which represent boundary-case granularities widely adopted in prior automated repair research. 
Although intermediate or hybrid granularities (\eg function-level or block-level generation) may offer additional trade-offs between contextual preservation and localized precision, we intentionally focus on these two canonical strategies to isolate their principal operational differences under complex, system-level build-repair scenarios. 
Future work may extend the benchmark to incorporate adaptive or hybrid repair strategies for finer-grained sensitivity analysis.

\textit{Conclusion Validity.}  
All models are evaluated under identical prompts, tool configurations, and iteration budgets to minimize bias. 
A potential threat to conclusion validity is the evolution of model APIs over time, as backend updates by providers can alter model behavior and affect the exact replicability of results. 
To mitigate this risk, we utilize the Model Context Protocol (MCP), a standardized and relatively stable interface for tool orchestration that remains robust against minor algorithmic shifts in underlying models. 
Furthermore, we include an open-source small language model (SLM), Qwen2.5-3B-Instruct, as a lightweight and fully reproducible baseline that can be deployed locally, thereby eliminating the uncertainties associated with closed-source API versioning. 
Disparities in backend infrastructure, such as API rate limits or transient model updates, may still introduce small fluctuations in timing or success statistics. 
To alleviate these biases and facilitate independent verification, we have released the complete dataset, prompt templates, and experimental scripts, allowing the community to replicate the reasoning trajectories and build outcomes under consistent settings.
To reduce random noise, results are aggregated across multiple packages to ensure that the reported metrics reflect stable performance trends rather than isolated instances.

\section{Conclusion}
We present \name{}, the first executable and architecture-aware benchmark designed to evaluate the ability of large language models to repair build failures in cross-ISA migration.
By combining real-world build environments, autonomous tool usage, and iterative feedback, the benchmark enables a comprehensive assessment of model reasoning, adaptation, and verification behaviors.
Through extensive experiments on six primary LLMs, together with an additional lightweight SLM baseline in the overall performance analysis, we find that iterative feedback substantially improves repair accuracy, yet long-log comprehension, procedural reasoning, and cross-file consistency remain significant challenges.
The analysis of tool invocation behaviors further reveals diverse strategies and levels of autonomy across models, highlighting the importance of structured tool orchestration in LLM-driven build repair.
Overall, \name{} fills an important benchmark gap by providing both a rigorous evaluation framework and a practical foundation for future research on LLM-based automation in software maintenance and system migration.
Future work will extend the benchmark to additional architectures and explore self-hosted build environments to strengthen reproducibility and long-term sustainability.

\begin{acks}
This work is supported by the National Natural Science Foundation of China (62272249, 62302244), and the Fundamental Research Funds for the Central Universities (XXX\-63253249).
\end{acks}

\bibliographystyle{ACM-Reference-Format}
\bibliography{main}

\appendix

\section{Prompt Templates}

To improve transparency and reproducibility, this appendix provides the complete prompt templates used in our experiments. 
All evaluated models are provided with identical prompts without model-specific tuning. 
The prompts are structured modularly, as described in Section~\ref{sub: prompt design}, and are programmatically updated at each iteration using execution feedback as detailed in Section~\ref{subsubsec: iterative reasoning}.

\subsection{Full-File Generation Prompt}

\begin{lstlisting}[style=promptstyle]
You are an advanced intelligent agent specializing in **end-to-end software build error resolution** when migrating from the x86_64 instruction set architecture to the aarch64 (ARM64) architecture.
Your mission is to **diagnose build failure logs** and then **apply the minimal, precise modifications** to ensure that the software package builds successfully on the aarch64 architecture.
You work across logs, specification files, and source code to ensure that the software package can be successfully rebuilt and verified.

## Guiding Principles
- **Minimalism**: Fix only the exact cause; no refactors or unrelated edits.
- **Completeness**: When modifying files, always output the **entire file content** (no partials/placeholders).
- **Preserve Structure**: Keep formatting, comments, and original style unless a specific line requires repair.
- **Syntactic Correctness**: Ensure all modified files compile/parse in their languages.

## Packaging & Stop Rules
- Work entirely within {temp_dir} (the package root).
- If you extracted the source, you MUST re-compress **the package root** before uploading.
- NEVER pass `"{temp_dir}/extracted"` to any tool; always use the package root `"{temp_dir}"`.
- The **final two tool calls of each attempt MUST be**:
  1) `compress_to_archive_tool[package_path="{temp_dir}"]` **(only if** an `extracted/` subdir exists or packaging is required)  
  2) `upload_file_to_obs_tool[package_path="{temp_dir}"]`
- If any of the final tools return an error, continue repairing and repeat the final sequence.

## Core Capabilities
**Automated Repair**
   - Work entirely within the temporary directory `{temp_dir}`
   - Apply the **minimal edits necessary** to resolve the issue
   - Preserve original archive format when repackaging (**only when necessary**)
   - Verify correctness by uploading to OBS (`upload_file_to_obs_tool`) and then check the build result using (`check_build_result`) tool. 
   - Note that the (`check_build_result`) involves actual build process in OBS, which **may take up to 10 minutes to complete**. 
   - Please be patient and wait for the result before proceeding. If the build fails, continue to analyze and repair until the build is successful or the maximum number of retries is reached.
   - Log all changes and save the final result to `{file_name}`


## Output format
For each modified file, output exactly in this format:

===FILE: relative/path/to/file===
<full file contents after modification>

## Important notes
- Do not output explanations, only the modified files in the above format.
- If multiple files are modified, repeat the `===FILE:...===` block for each.
- If the build fails, analyze the failure reason and continue to repair. Repeat the repair steps until the build is successful or the maximum number of retries is reached.

\end{lstlisting}

\subsection{Patch-Generation Prompt}
\begin{lstlisting}[style=promptstyle]
You are an advanced intelligent agent specializing in **end-to-end software build error resolution** when migrating from the x86_64 instruction set architecture to the aarch64 (ARM64) architecture.
Your mission is to **diagnose build failure logs** and then **apply the minimal, precise modifications** so that the package builds successfully on the target architecture.
You work across logs, specification files, and source code to ensure that the software package can be successfully rebuilt and verified.

## Guiding Principles
- **Minimalism**: Fix only the exact cause; no refactors or unrelated edits.
- **Preserve Structure**: Keep formatting, comments, and original style unless a specific line requires repair.
- **Syntactic Correctness**: Ensure all modified files compile/parse in their languages.

## Core Capabilities
**Automated Repair**
   - Work entirely within the temporary directory `{temp_dir}`
   - Apply the **minimal edits necessary** to resolve the issue
   - Preserve original archive format when repackaging (**only when necessary**)
   - Verify correctness by uploading to OBS (`upload_file_to_obs_tool`) and then check the build result using (`check_build_result`) tool. 
   - Note that the (`check_build_result`) involves actual build process in OBS, which **may take up to 10 minutes to complete**. 
   - Please be patient and wait for the result before proceeding. If the build fails, continue to analyze and repair until the build is successful or the maximum number of retries is reached.
   - Log all changes and save the final result to `{file_name}`

## Patch format
- You MUST generate a **standard git-style unified diff** (as seen in GitHub PRs).
- The patch may modify multiple files; each file must include headers and hunks:
  - File header lines:
    diff --git a/<relpath> b/<relpath>
    --- a/<relpath>        (or --- /dev/null for new files)
     b/<relpath>        (or  /dev/null for deleted files)
  - One or more hunk headers:
    @@ -<start>[,<len>] <start>[,<len>] @@[ optional title ]
  - Hunk body lines starting with:
    ' ' (context), '+' (added), '-' (deleted)
- Critical Warnings (avoid errors):
  1) **Bad hunk header error**: Hunk header CANNOT be write as `@@` (must include line ranges). 
      - Wrong: `@@` (causes "Bad hunk header")
      - Correct: `@@ -41,11 41,16 @@` (explicit start/len)
  2) **Hunk failed error**: Tool uses **strict context matching** (no fuzzy apply). 
      - Ensure context lines (starting with ' ') in patch are EXACTLY the same as target file.
      - Include 3-5 context lines around changes (not too few) to avoid mismatch.
      - Verify line numbers in hunk header match the actual target file (e.g., `-41,11` means start at line 41, 11 lines total).
- **Paths must be relative to the repository root** (i.e., {temp_dir}). Do NOT use absolute paths.
- Always include enough context lines to ensure the hunk applies strictly.
- Do NOT emit pseudo formats like "*** Begin Patch" or hunks without line ranges.

## Packaging & Stop Rules
- Work entirely within {temp_dir} (the package root).
- If you extracted the source, you MUST re-compress **the package root** before uploading.
- NEVER pass `"{temp_dir}/extracted"` to any tool; always use the package root `"{temp_dir}"`.
- To apply your patch, call: apply_git_unified_patch_tool(repo_root="{temp_dir}", patch_text="<FULL PATCH TEXT>")
- The **final two tool calls of each attempt MUST be**:
  (1) `compress_to_archive_tool[package_path="{temp_dir}"]` **(only if** an `extracted/` subdir exists or packaging is required)  
  (2) `upload_file_to_obs_tool[package_path="{temp_dir}"]`
- If any of the final tools return an error, continue repairing and repeat the final sequence.

## Important Notes
- **Do not print full file contents** or handcrafted patch blocks in the chat.
- Only use the tools above to make minimal changes, then proceed to package and verify.
- If the build fails, analyze, minimally edit again (via unified diff), repackage if needed, upload, and re-check until success or retries are exhausted.


\end{lstlisting}

\label{append1}

\section{Granular Success Rate Analysis by Failure Subcategories}

To provide a more fine-grained view of model behavior across different failure types, Table~\ref{tab:failure_subcategory_success} reports the repair success rates of GPT-5 further stratified by failure subcategories. 

The subcategory-level breakdown reveals several additional insights into the repair behavior of GPT-5 across different failure types. 
First, certain configuration-related failures appear relatively amenable to automated repair. In particular, subcategories such as \textit{Compiler and Flag Configuration Errors} exhibit consistently high success rates across both repair modes and migration directions. This suggests that many configuration-level issues can be effectively resolved through modifications to build specifications or compiler flags.
Second, packaging-related failures show a notable advantage for \textit{Patch Generation} in several cases. For example, failures involving RPM scripts or missing artifacts often benefit from localized edits that directly target problematic build steps. This observation indicates that fine-grained patching can be particularly effective when the repair requires small procedural adjustments rather than broader structural changes.
Third, failures related to deeper code semantics, such as \textit{Compiler and Type System Errors}, tend to exhibit larger performance differences between repair strategies. In these cases, \textit{Full File Generation} occasionally achieves higher success rates, suggesting that regenerating the entire file may help the model maintain better global consistency when resolving complex semantic issues.
Finally, the results also reflect the inherent asymmetry between the two migration directions. In several subcategories, the repair success rate is noticeably lower when migrating toward \textit{x86\_64}, highlighting the additional challenges involved in reverse ISA migration.

Overall, these fine-grained statistics complement the category-level analysis presented in the main text and illustrate that the effectiveness of different repair strategies varies substantially depending on the underlying failure semantics.

\begin{table*}[t]
\centering
\caption{
Repair success rate (\%) of GPT-5 across failure categories and their subcategories under the two repair modes (Full File Generation and Patch Generation). 
}
\label{tab:failure_subcategory_success}
\newcolumntype{L}[1]{>{\raggedright\arraybackslash}p{#1}}
\newcolumntype{C}[1]{>{\centering\arraybackslash}p{#1}}
\setlength{\tabcolsep}{4pt}
\renewcommand{\arraystretch}{1.15}

\begin{tabularx}{\textwidth}{
  L{0.18\textwidth}
  X
  C{0.085\textwidth}
  C{0.085\textwidth}
  C{0.085\textwidth}
  C{0.085\textwidth}
}
\toprule
\multirow{2}{*}{\textbf{Category}} & 
\multirow{2}{*}{\textbf{Subcategory}} &
\multicolumn{2}{c}{\textbf{x86\_64 $\rightarrow$ aarch64}} &
\multicolumn{2}{c}{\textbf{aarch64 $\rightarrow$ x86\_64}} \\
\cmidrule(lr){3-4}\cmidrule(lr){5-6}
 & & \textbf{Full} & \textbf{Patch} & \textbf{Full} & \textbf{Patch} \\
\midrule

\multirow{2}{*}{Preparation Error}
& Environment and Dependency Issues
& \textbf{46.67} & 40.00 & \textbf{33.33} & 27.78 \\
& Compiler and Flag Configuration Errors
& 80.95 & \textbf{85.71} & 66.67 & 66.67 \\

\midrule

\multirow{3}{*}{Compilation Error}
& Build/Compiler Configuration Failures
& \textbf{72.73} & 69.70 & 33.33 & 33.33 \\
& Compiler and Type System Errors
& \textbf{53.57} & 28.57 & 13.33 & \textbf{33.33} \\
& Warning Escalation and Policy Failures
& \textbf{50.00} & 33.33 & 14.29 & \textbf{28.57} \\

\midrule

\multirow{3}{*}{Packaging Error}
& Missing or Unpackaged Artifacts
& 44.44 & \textbf{55.56} & \textbf{60.00} & 40.00 \\
& RPM Script and Build Step Failures
& 25.00 & \textbf{50.00} & 25.00 & \textbf{75.00} \\
& Specification or Policy Violations
& 100.00 & 100.00 & 100.00 & 100.00 \\

\midrule

\multirow{3}{*}{Test Failure}
& Functional and Assertion Failures
& 50.00 & 50.00 & 38.46 & \textbf{69.23} \\
& Environment Setup Failures
& \textbf{66.67} & 50.00 & 0.00 & \textbf{33.33} \\
& Runtime and Execution Failures
& 40.00 & 40.00 & \textbf{50.00} & 33.33 \\

\midrule

Env/Infra Error
& Host or Virtualization Failure
& \textbf{60.00} & 40.00 & 33.33 & 33.33 \\

\bottomrule
\end{tabularx}
\end{table*}
\label{append2}








\end{document}